\documentclass[final,3p,times]{elsarticle}
\usepackage{amssymb}
\usepackage{comment}
 \usepackage{amsmath}
 \usepackage{booktabs}
\usepackage{multicol}
\usepackage{multirow}
\usepackage{placeins}
\usepackage{subcaption}
\usepackage{stfloats}
\usepackage{float}
\usepackage{color,soul}
\usepackage{xcolor}
\setstcolor{red}
\usepackage[colorlinks=true,
            linkcolor=blue]{hyperref}
\usepackage[font=normalsize,skip=3pt]{caption} 
\journal{Computers in Biology and Medicine}

\begin{document}

\begin{frontmatter}

%% Title, authors and addresses

%% use the tnoteref command within \title for footnotes;
%% use the tnotetext command for theassociated footnote;
%% use the fnref command within \author or \address for footnotes;
%% use the fntext command for theassociated footnote;
%% use the corref command within \author for corresponding author footnotes;
%% use the cortext command for theassociated footnote;
%% use the ead command for the email address,
%% and the form \ead[url] for the home page:
%% \title{Title\tnoteref{label1}}
%% \tnotetext[label1]{}
%% \author{Name\corref{cor1}\fnref{label2}}
%% \ead{email address}
%% \ead[url]{home page}
%% \fntext[label2]{}
%% \cortext[cor1]{}
%% \affiliation{organization={},
%%             addressline={},
%%             city={},
%%             postcode={},
%%             state={},
%%             country={}}
%% \fntext[label3]{}

\title{Potential and challenges of generative adversarial networks \\for super-resolution in 4D Flow MRI}

% Authors
\author[inst1]{Oliver Welin Odeback\corref{cor1}}
\ead{oliver.welin.odeback@ki.se}
\author[inst2]{Arivazhagan Geetha Balasubramanian}
\author[inst3]{Jonas Schollenberger}
\author[inst4,inst5]{Edward Ferdian}
\author[inst5,inst6]{Alistair A. Young}
\author[inst7]{C. Alberto Figueroa}
\author[inst8]{Susanne Schnell}
\author[inst2]{Outi Tammisola}
\author[inst9]{Ricardo Vinuesa}
\author[inst1,inst10]{Tobias Granberg}
\author[inst1,inst11]{Alexander Fyrdahl}
\author[inst1,inst12]{David Marlevi}

\cortext[cor1]{Corresponding author.}

% Affiliations
\affiliation[inst1]{organization={Dept. Molecular Medicine and Surgery, Karolinska Institutet},%Department and Organization
            addressline={Karolinska Universitetssjukhuset Solna (L1:00)}, 
            city={Stockholm},
            postcode={171 76},
            country={Sweden}}
\affiliation[inst2]{organization={FLOW, Engineering Mechanics, KTH Royal Institute of Technology},%Department and Organization
            addressline={Osquars Backe 18}, 
            city={Stockholm},
            postcode={100 44}, 
            country={Sweden}}
\affiliation[inst3]{organization={Department of Radiology and Biomedical Imaging, University of California San Francisco},%Department and Organization
            addressline={505 Parnassus Avenue}, 
            city={San Francisco},
            postcode={94143}, 
            state={CA},
            country={USA}}
\affiliation[inst4]{organization={Faculty of Informatics, Telkom University},%Department and Organization
            addressline={Jl.Telekomunikasi No. 1, Terusan Buahbatu}, 
            city={Bandung},
            postcode={40257}, 
            state={West Java},
            country={Indonesia}}
\affiliation[inst5]{organization={Auckland Bioengineering Institute, University of Auckland},%Department and Organization
            addressline={Bioengineering House, 70 Symonds St}, 
            city={Grafton},
            postcode={1010},
            country={New Zealand}}
\affiliation[inst6]{organization={School of Biomedical Engineering \& Imaging Sciences, King's College London},%Department and Organization
            addressline={1 Lambeth Palace Rd, South Bank}, 
            city={London},
            postcode={SE1 7EU},
            country={UK}}
\affiliation[inst7]{organization={Department of Biomedical Engineering, University of Michigan},%Department and Organization
            addressline={1107 Carl A. Gerstacker Bldg
            2200 Bonisteel Blvd.}, 
            city={Ann Arbor},
            postcode={48109-2099},
            state={MI},
            country={USA}}
\affiliation[inst8]{organization={Department of Physics, University of Greifswald},%Department and Organization
            addressline={Felix-Hausdorff-Str. 6}, 
            city={Greifswald},
            postcode={174 89},
            country={Germany}}
\affiliation[inst9]{organization={Department of Aerospace Engineering, University of Michigan},%Department and Organization
            addressline={1320 Beal Avenue}, 
            city={Ann Arbor},
            postcode={48109-2140},
            state={MI},
            country={USA}}
\affiliation[inst10]{organization={Department of Neuroradiology, Karolinska University Hospital},%Department and Organization
            addressline={Hälsovägen 13, O42}, 
            city={Stockholm},
            postcode={141 86},
            country={Sweden}}
\affiliation[inst11]{organization={Department of Clinical Physiology, Karolinska University Hospital},%Department and Organization
            addressline={Eugeniavägen 3, A8:01}, 
            city={Solna},
            postcode={171 64},
            country={Sweden}}
\affiliation[inst12]{organization={Institute for Medical Engineering and Science, Massachusetts Institute of Technology},%Department and Organization
            addressline={45 Carleton St}, 
            city={Cambridge},
            postcode={02142},
            state={MA},
            country={USA}}

\begin{abstract}

Time-resolved three-dimensional phase-contrast MRI (4D Flow MRI) enables non-invasive quantification of blood flow and derivation of hemodynamic parameters. However, its clinical application is limited by low spatial resolution and noise, particularly affecting velocity measurements near vessel walls. Machine learning-based super-resolution has shown promise in addressing these limitations, but challenges remain, not least in recovering near-wall velocities. Generative adversarial networks (GANs) offer a compelling solution, having demonstrated strong capabilities in restoring sharp boundaries in non-medical super-resolution settings. Yet, their application in 4D Flow MRI remains unexplored, with implementation challenged by known issues such as training instability and non-convergence. In this study, we investigate GAN-based super-resolution and denoising in 4D Flow MRI. Training and validation were conducted using patient-specific cerebrovascular \textit{in-silico} models, converted into synthetic images via an MR-true reconstruction pipeline, with complementary validation on \textit{in-vivo} acquisitions. A dedicated GAN architecture was implemented and evaluated across three adversarial loss functions: Vanilla, Relativistic, and Wasserstein. Our results demonstrate that the proposed GAN improved near-wall velocity recovery compared to a non-adversarial reference (vector Normalized Root Mean Square Error (vNRMSE): 6.9\% vs. 9.6\%); however, that implementation specifics are critical for stable network training. While Vanilla and Relativistic GANs proved unstable compared to generator-only training (vNRMSE: 8.1\% and 7.8\% vs. 7.2\%), a Wasserstein GAN demonstrated optimal stability and incremental improvement (vNRMSE: 6.9\% vs. 7.2\%). Moreover, strong \textit{in-vivo} performance supports clinical translation. Together, these findings highlight the potential of GAN-based super-resolution in enhancing 4D Flow MRI, particularly in challenging cerebrovascular regions, while emphasizing the importance of carefully selecting adversarial training strategies.

\end{abstract}

\begin{keyword}
%% keywords here, in the form: keyword \sep keyword
4D Flow MRI \sep Generative adversarial networks \sep Cerebrovascular flow \sep Deep learning \sep Super-resolution
%% PACS codes here, in the form: \PACS code \sep code
%\PACS 0000 \sep 1111
%% MSC codes here, in the form: \MSC code \sep code
%% or \MSC[2008] code \sep code (2000 is the default)
%\MSC 0000 \sep 1111
\end{keyword}

\end{frontmatter}

%% \linenumbers

\section{Introduction}

Cardiovascular diseases remain a leading cause of morbidity and mortality worldwide, driving the need for improved diagnostic and prognostic tools. Among available imaging modalities, time-resolved three-dimensional phase-contrast MRI (\textit{4D Flow MRI}) provides unique capabilities for non-invasive assessment of blood flow, but its clinical use is limited by a fundamental trade-off between spatial resolution, acquisition time, and image noise. Overcoming these limitations is critical for reliable hemodynamic quantification - particularly in the narrow and tortuous cerebrovasculature, where accurate estimation of near-wall flow and derived biomarkers such as wall shear stress remains especially challenging. Herein, recent advances in deep learning, and in particular GANs, offer an opportunity to enhance 4D Flow MRI retrospectively, yet their feasibility and utility in this domain remain unexplored.

The aim of this study is therefore to investigate the potential and limitations of GAN-based models for super-resolving and denoising 4D Flow MRI, with a particular emphasis on enhancing flow recovery at vessel boundaries. To address this, we present a dedicated core GAN setup for velocity field super-resolution, which we then systematically evaluate across three adversarial loss formulations, all in order to outline the potential as well as the challenges of GAN training stability and reconstruction accuracy in a 4D Flow MRI setting. The networks were trained and validated on synthetic cerebrovascular 4D Flow MRI data originating from patient-specific CFD models, with additional validation on \textit{in-vivo} acquisitions. The complex and narrow geometry of the intracranial vasculature provides a particularly suitable setting for evaluating boundary-focused super-resolution. The complete training setup and models are publicly available at \href{https://github.com/oliverwelinodeback/4DFlowGAN}{https://github.com/oliverwelinodeback/4DFlowGAN}.

%The networks were trained and validated on synthetic cerebrovascular 4D Flow MRI data derived from patient-specific CFD models, complemented by in-vivo acquisitions from a healthy volunteer and a patient with multiple sclerosis. The complex and narrow geometry of the intracranial vasculature provides a particularly suitable setting for evaluating boundary-focused super-resolution.

\section{Background and Related Work}

Single image super-resolution (SISR) is the task of reconstructing a high-resolution (HR) image from its low-resolution (LR) counterpart, recovering features blurred or not fully conveyed in the input LR data. Driven by its relevance in domains spanning from video enhancement to clinical imaging, SISR has gained considerable attention within the computer vision community~\cite{kappeler2016video, wang2020deep, li2021review}. While traditional deterministic methods have been extensively studied~\cite{hou1978cubic, keys2003cubic, park2003super}, deep learning approaches - particularly those based on convolutional neural networks (CNNs) - have become the dominant strategy owing to their ability to learn complex spatial mappings directly from data~\cite{park2003super, dong2015image, kim2016accurate, tong2017image, ledig2017srgan, sajjadi2017enhancenet, wang2018esrgan}. In recent years, both network architectures and training strategies have evolved significantly in this domain~\cite{kim2016accurate, tong2017image, wang2018esrgan, karnewar2020msg, liang2022details, lu2022transformer}, consistently improving super-resolution performance.
 
As part of this progressive SISR development, generative adversarial networks (GANs)~\cite{goodfellow2014gan} have attracted specific attention by enabling the reconstruction of visually sharper images with enhanced contrast and boundary recovery~\cite{ledig2017srgan, sajjadi2017enhancenet, wang2018esrgan}. Herein, GANs consist of two CNN-based networks, a generator and a discriminator, where the generator produces synthetic data while the discriminator tries to distinguish these from real data. Through so called adversarial training, the discriminator then forces the generator to produce predictions that become increasingly indistinguishable from the real, reference data. A key milestone in GAN-based SISR was SRGAN~\cite{ledig2017srgan}, demonstrating how GANs can significantly improve perceptual quality compared to conventional fully connected models. Subsequent work~\cite{sajjadi2017enhancenet, wang2018esrgan, karnewar2020msg} has achieved improved texture fidelity and perceptual quality through architectural refinements and alternative adversarial loss formulations. For instance, the so called enhanced SRGAN (ESRGAN)~\cite{wang2018esrgan} extended SRGAN by adopting a relativistic discriminator~\cite{jolicoeur2018relativistic}, leading to further gains in perceptual sharpness and SR recovery.

Among a variety of application areas, SR techniques have been applied to reconstruct flow fields from sparse data in fluid mechanics~\cite{vinuesa2023transformative}, and have also gained increasing attention in the medical imaging domain, where they have demonstrated the ability to enhance resolution and reduce image noise across a variety of imaging modalities~\cite{pham2019multiscale, park2018computed, christensen2020super}. A prominent application example is 4D Flow MRI~\cite{markl20124d}; a modality enabling non-invasive full-field quantification of blood flow velocity, but where direct trade-offs between image resolution, acquisition time, and effective image noise  limit clinical applicability~\cite{stankovic2014reproducibility}. In particular, when targeting narrow vascular compartments such as the portal, fetal, or intracranial settings, resolution-based biases have been reported with direct impact on accurate hemodynamic quantification at typical clinical resolutions~\cite{aristova2019standardized, marlevi2021noninvasive, hyodo20224d}. While higher resolution can be achieved through dedicated acquisition protocols, novel sampling patterns, or high-field systems~\cite{schnell2016four, Aristova2022, gottwald2020high}, these approaches are constrained by system availability. In contrast, deep-learning based SISR opens up for \textit{post-scan} resolution enhancement, and recovery of features otherwise unattainable in typical clinical acquisition setting~\cite{fathi2020super, Ferdian20204DFlowNet, Ferdian2023Cerebrovascular, shit2022srflow, saitta2024implicit}. In this context, the use of residual CNNs has been the dominant architectural approach, with notable examples including Rutkowski et al.~\cite{Rutkowski2021}, 4DFlowNet~\cite{Ferdian20204DFlowNet, Ferdian2023Cerebrovascular} and SRFlow~\cite{shit2022srflow}. However, while these models have demonstrated success in various vascular domains, reconstruction of near-wall velocities remains challenging, where steep velocity gradients and partial voluming effects pose persistent obstacles for accurate network recovery~\cite{Rutkowski2021, Ferdian2023Cerebrovascular}. This also represents a clinical obstacle, where biomarkers including wall shear stress (WSS), or coupled endothelial activation, rely on accurate near-wall velocity capture~\cite{szajer2018comparison}. 

In this setting, the noted GAN-based developments carry particular promise due to their ability to recover high-frequency details from low-resolution and noisy input, with demonstrated feasibility of reconstructing velocity fields from limited or sparse data in fluid mechanics~\cite{guemes2021coarse, yousif2023deep}, making them strong candidates for boundary-focused enhancement in 4D Flow MRI~\cite{ledig2017srgan, sajjadi2017enhancenet, wang2018esrgan}. Nevertheless, while tested across other imaging modalities~\cite{you2019ct, zhang2022soup}, translation of SISR GANs into a 4D Flow MRI domain remains unexplored. A specific obstacle persists in GANs being characterized by training difficulties, with several works reporting issues related to training instability, mode collapse and non-convergence, all in part relating to the non-convex nature of the optimization setup~\cite{saxena2021review, gui2021review, arjovsky2017wasserstein, karnewar2020msg, mescheder2018training}. Further, reports of GAN-based models being prone to high-frequency artifacts compromising reconstruction accuracy may be seen as particularly problematic in clinical settings, where accurate and reliable precision is critical~\cite{blau2018perception, zhang2018unreasonable, liang2022details}. 

The challenges of training GANs have motivated ongoing efforts aimed at stabilizing their training dynamics and refining the adversarial loss functions. These include architectural innovations, such as MSG-GAN~\cite{karnewar2020msg}, which propagates gradients across multiple resolutions to improve multi-scale coherence, and loss function reformulations like the Wasserstein GAN (WGAN)~\cite{arjovsky2017wasserstein}, which replaces the original adversarial loss with the Earth Mover (Wasserstein-1) distance to ensure smoother and more informative gradient flow. While such advancements have shown promise across certain imaging tasks, no clear consensus has emerged regarding which strategies are most effective under specific data characteristics and application domains. In particular, the behavior and possible utility of GAN-based models in the context of 4D Flow MRI remain unexplored.

\section{Methods}

\subsection{\texorpdfstring{Training and validation data}{Training and validation data}}

\begin{figure*}[b!]
  \centering
  \includegraphics[width=\linewidth]{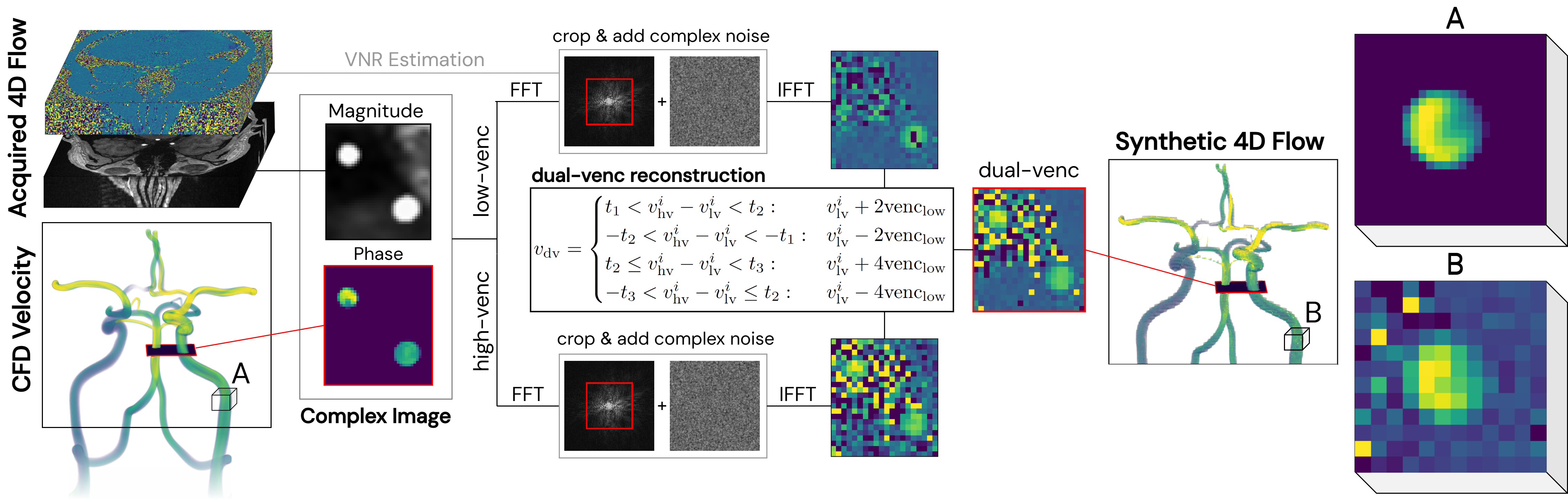}
  \caption{Overview of the dual-venc reconstruction and synthetic 4D Flow MRI generation process.}
  \label{fig:DV-recon}
\end{figure*}

Training supervised super-resolution networks relies on having matched low- and high-resolution datasets as input. In the setting of super-resolution 4D Flow MRI, these datasets would ideally originate from paired clinical acquisitions from the very same subject. However, in reality such acquisitions are difficult to achieve, with paired acquisitions suffering from inter-scan variabilities (patient movements; cardiorespiratory variations; etc.), and since high-resolution scans are affected by inherent noise, they are unsuitable as noise-free references. Instead, to serve as an idealized testbed for supervised super-resolution network development, and similar to prior work in the area \cite{Ferdian20204DFlowNet, Ferdian2023Cerebrovascular, Rutkowski2021, shit2022srflow, dirix2022synthesis}, we utilize synthetically generated 4D Flow MRI data, originating from patient-specific computational fluid dynamics (CFD) simulations and with realistic imaging features generated by MRI-based data downsampling (see Section~\ref{sec:synth_4Dflow}). Importantly, this strategy provides perfectly matched low- and high-resolution datasets, including idealized noise-free high-resolution training data, together enabling a robust foundation for super-resolution model training and evaluation.

\subsubsection{Patient specific in-silico models}

With a specific focus on intracranial applications, patient-specific CFD models of the arterial cerebrovasculature served as the foundational input for generating training and validation data. With modeling details outlined in \cite{schollenberger2021}, and with identical models used in previous work for super-resolution imaging \cite{Ferdian2023Cerebrovascular, ericsson2024generalized}, below follows a brief recapitulation of key modeling concepts:

Computational models were developed by combining multiple patient-specific MR-acquired image sets (T1-weighted, time-of-flight (TOF), 2D phase-contrast (PC), and arterial spin labeling (ASL), together providing anatomical and functional model boundary condition inputs~\cite{schollenberger2021}. With models defined and meshed, we performed CFD simulations using the open-source CRIMSON framework~\cite{arthurs2021crimson} to solve the imposed incompressible Navier–Stokes equations. Nodal velocity data were extracted once periodic flow was reached (empirically determined after $\leq$ 4 cardiac cycles) with this data subsequently used as input for generating synthetic reference 4D Flow MRI data (see Section~\ref{sec:synth_4Dflow}). 

As in the work by Ferdian et al.~\cite{Ferdian2023Cerebrovascular}, data from four different image sets were used to generate four different models with varying degrees of cerebrovascular disease: one healthy reference case (subject 1); one with severe stenosis in the right proximal internal carotid artery (ICA) (subject 2); one with bilateral ICA stenoses (subject 3a); and one in the bilateral ICA case \textit{after} surgical re-opening of the right ICA stenosis (subject 3b). Further modeling details along with validation against vessel-specific ASL measurements can be found in previous publications~\cite{schollenberger2021, schollenberger2023magnetic, Ferdian2023Cerebrovascular}.

\subsubsection{Synthetic 4D Flow MRI data}
\label{sec:synth_4Dflow}

With output data generated as outlined above, the field-of-view for the synthetic image generation was selected to cover the intracranial vessels (see Figure~\ref{fig:DV-recon}). Nodal output was then mapped onto uniformly voxelized image grids with isotropic resolutions of 0.75, 0.5, and 0.375~mm, and a time discretization of 10~ms, seeking to capture multiple spatial scales to allow for a sufficient amount of training data.  

To convert these noise-free, high-resolution velocity data into clinically relevant, synthetic low-resolution 4D Flow MRI equivalents, a dedicated acquisition-mimicking pipeline was implemented. Specifically, dual velocity encoding (dual-venc)~\cite{Schnell2017, Aristova2022} is often employed to improve the VNR of clinical 4D Flow acquisitions without introducing aliased data. Incorporating \textit{in-vivo} magnitude images and a dedicated dual-venc reconstruction procedure, the pipeline was created to closely imitate a realistic clinical acquisition relevant for intracranial applications. The following steps outline the complete pipeline, with an illustrative overview given in Figure~\ref{fig:DV-recon}:

\begin{enumerate}

\item \textbf{Complex value signal:} To enable the creation of a complex value input signal, the synthetic velocity data $\mathbf{v}_{\text{HR}}$ of the CFD had to be paired with a corresponding magnitude image $m$. For this, magnitude images from clinically acquired 4D Flow MRI data at 3T were used (Siemens Magnetom Skyra, prospective \textit{k-t} GRAPPA dual-venc (120/60 cm/s) acquisition), with images retrospectively assembled from five individuals of an IRB-approved study including informed consent. To align clinical and synthetic image data at different spatial resolutions, a 3D interpolation was performed to align the transverse plane resolution of the synthetic velocity data to the reference magnitude data. Transverse images were then stacked vertically, accounting for remaining out-of-plane resolution discrepancies to create two complete volume sets at matching spatial resolution. To account for vessel and non-vessel magnitude differences, voxel-wise overlap between the CFD and the \textit{in-vivo} magnitude vessel masks was evaluated. Specifically, where overlap existed the corresponding in-vivo magnitude value was assigned; for non-overlapping voxels, the mean vessel or non-vessel \textit{in-vivo} intensity was assigned, depending on the CFD region. This yielded magnitude volumes $m$ that were fully co-registered to the CFD velocity fields while preserving realistic intensity variations. Additionally, to emulate realistic noise characteristics observed in clinical acquisitions, magnitude intensities below an empirically determined threshold of 30 were set to zero.

\item \textbf{Phase image creation:} With magnitude and velocity images aligned, we transformed the simulated velocity field \(v\) into corresponding phase images $\phi$ by:
\begin{equation}
\phi = \left(\frac{\mathbf{v}_{\text{HR}}}{\text{venc}}\right)\,\pi,
\end{equation}
where venc denotes the user-defined synthetic velocity encoding parameter. With this, complex image sets could then be created by combining the aligned magnitude volume \(m\) with the phase data $\phi$ by:
\begin{equation}
\mathbf{v}_{\text{complex}} = m \cdot e^{i\,\phi}.
\end{equation}
To account for a dual velocity encoded representation, 
both low \(\bigl(\text{venc}_{\text{low}} = 0.5 \text{--} 0.7 \,\text{m/s}\bigr)\) and high venc \(\bigl(\text{venc}_{\text{high}} = 2\times\text{venc}_{\text{low}}\) or \( \text{max}(\mathbf{v}_t) \) if \( \max(\mathbf{v}_t) \geq 2\times\text{venc}_{\text{low}} \text{ at timestep $t$}\bigr)\) images were generated, yielding two corresponding complex image sets, respectively, \(\mathbf{v}_{\text{lv}}\) and \(\mathbf{v}_{\text{hv}}\).

\item \textbf{k-space cropping and noise:} Low- and high-venc complex images were transferred into k-space using a Fast Fourier Transform (FFT). With high-resolution data in k-space, low-resolution equivalents were generated by centrally cropping k-space using rectangular truncation without additional windowing, decreasing the spatial resolution by a factor of 2. This was performed across both high and low-venc channels. Complex zero-mean Gaussian noise was then added with standard deviation, \(\sigma\), determined according to a user-defined target signal-to-noise ratio (TSNR), following the approach of Ferdian et al.~\cite{Ferdian2023Cerebrovascular}. For a given time step, either a high SNR (TSNR = 8--12) or a low SNR (TSNR = 2--6) was randomly assigned, matching clinically observed levels in similar-resolution dual-venc acquisitions while allowing the network to train across various noise conditions. Note that for high-venc images, we exclusively applied noise at TSNR = 15 to account for the increased Gibbs ringing artifacts observed which are associated with larger venc values \cite{parker1987gibbs}.

\item \textbf{Dual-venc reconstruction:} k-space data were returned to image space via an inverse FFT algorithm and then separated into velocity and magnitude images, using the same low- and high-venc values selected in step 2. A modified dual-venc reconstruction algorithm~\cite{Schnell2017, Aristova2022}  was applied to correct for aliasing in the low-venc velocity images, allowing for up to six phase wraps. The unwrapped dual-venc velocity $\mathbf{v}_{\text{dv}}$ was computed as: 
\end{enumerate}
\begin{equation}
\mathbf{v}_{\text{dv}} = 
\begin{cases}
t_1 < \mathbf{v}_{\text{hv}}^i - \mathbf{v}_{\text{lv}}^i < t_2 :  & \mathbf{v}_{\text{lv}}^i + 2\text{venc}_{\text{low}} \\
-t_2 < \mathbf{v}_{\text{hv}}^i - \mathbf{v}_{\text{lv}}^i < -t_1 :  & \mathbf{v}_{\text{lv}}^i - 2\text{venc}_{\text{low}} \\
t_2 \leq \mathbf{v}_{\text{hv}}^i - \mathbf{v}_{\text{lv}}^i < t_3 :  & \mathbf{v}_{\text{lv}}^i + 4\text{venc}_{\text{low}} \\
-t_3 < \mathbf{v}_{\text{hv}}^i - \mathbf{v}_{\text{lv}}^i \leq -t_2 :  & \mathbf{v}_{\text{lv}}^i - 4\text{venc}_{\text{low}} \\
t_3 \leq \mathbf{v}_{\text{hv}}^i - \mathbf{v}_{\text{lv}}^i < t_4 :  & \mathbf{v}_{\text{lv}}^i + 6\text{venc}_{\text{low}} \\
-t_4 < \mathbf{v}_{\text{hv}}^i - \mathbf{v}_{\text{lv}}^i \leq -t_3 :  & \mathbf{v}_{\text{lv}}^i - 6\text{venc}_{\text{low}}
\end{cases}
\end{equation}
\begin{enumerate}
\item[] where the thresholds were defined as $t_1 = 1.2\text{venc}_{\text{low}}$, $t_2 = 3\text{venc}_{\text{low}}$, $t_3 = 5\text{venc}_{\text{low}}$, and $t_4 = 7\text{venc}_{\text{low}}$. The resulting $\mathbf{v}_{\text{dv}}=\mathbf{v}_{\text{LR}}$ served as the synthetic low-resolution 4D Flow MRI velocity field. 
 \end{enumerate}

\subsubsection{Patch extraction and data augmentation}
\label{sec:patch_extr}
To generate a larger number of training sets, the input high- and low-resolution data were split into several smaller patches on which the network was trained. This patch-based strategy not only augmented our training set given the limited number of input models (\(n=4\)), but also minimized the influence of global geometry, enabling the network to focus on local hemodynamic patterns across various vascular sections. 

To generate patches, input models were divided into train ($n=2$, subject 1 and 2), validation ($n=1$, subject 3a), and test ($n=1$, subject 3b) sets. For the training and validation sets, cubic patches \(\mathbf{x}_{\text{HR}}\) of size \(24^3\) voxels were randomly extracted from the high-resolution velocity data \(\mathbf{v}_{\text{HR}}\), with each voxel containing three velocity components. To ensure that each patch contains flow, while still allowing the smallest vessels to be sampled, only patches with at least 5\% fluid-containing voxels were retained. Due to the relatively high surface-to-volume ratio, this sampling strategy results in substantial representation of near-wall regions within each patch. When identified via binary erosion of the vessel mask, boundary voxels constituted approximately 38\% of all fluid voxels across the test set. Corresponding patches \(\mathbf{x}_{\text{LR}}\) of size \(12^3\) voxels were then extracted from the low-resolution data $\mathbf{v}_{\text{LR}}$ at the same spatial location, creating aligned patch pairs of high and low resolution. 

To mitigate directional biases, each patch pair was then augmented by applying rigid rotations of 90°, 180°, and 270° in all Cartesian directions, respectively. As such, a total of 69,600 training and 1,600 validation patch-pairs $(\mathbf{x}_{\text{HR}}, \mathbf{x}_{\text{LR}})$ were created for network training and model selection.

For testing or general data inference, input data $\mathbf{v}_{\text{LR}}$ was also cropped into similar patch-like structures. Specifically, low-resolution patches $\mathbf{x}_{\text{LR}}$ were extracted using a stride of 8 in all directions, yielding a two-voxel overlap between neighboring patches. After prediction, only non-overlapping regions were retained to form the final super-resolved output $\mathbf{v}_{\text{SR}}$.

\subsubsection{In-vivo acquisitions and post-processing}
\label{sec:invivo_data}

In addition to the synthetic datasets, evaluation was also performed on \textit{in-vivo} 4D Flow MRI acquisitions in order to validate direct clinical translation. Imaging was performed on one healthy volunteer (male, 28 years) and one patient with neuroinflammatory disease (multiple sclerosis, male, 39 years), following institutional review board (IRB) approval and informed consent. Both scans were acquired at 3T (Siemens MAGNETOM Skyra, Erlangen, Germany) using a 20-channel head/neck coil. The protocol comprised a high-resolution dual-venc 4D Flow MRI sequence (low/high-venc = 60/120 cm/s) with k-t GRAPPA acceleration (R = 5) and prospective gating, with a ROI centered on the Circle of Willis. In both cases, flow imaging was performed at an isotropic spatial resolution of 0.6~mm (matrix size: 280 × 320 × 80; FOV: 160 × 108 × 48~mm), with a temporal resolution of 103.6~ms (TR/TE = 103.6/4.96~ms, flip angle 15°). The acquisition time was approximately 40 minutes. Although such high-resolution protocols are impractical for routine clinical use, they provide a valuable reference for evaluation. Both sets were corrected for eddy currents. Additionally, a 3D time-of-flight gradient echo sequence (TR/TE = 21.11/3.69~ms, flip angle 20°) was acquired for segmentation. Specifically, a semi-automatic MATLAB tool~\cite{Schnell2017} was used to co-register the TOF images with the 4D Flow MRI data via rigid registration, implemented using functions from the SPM12 toolbox (Statistical Parametric Mapping, version 12)~\cite{ashburner2005unified}.

To generate paired, clinically feasible data free from inter-scan variability, we used the downsampling procedure described in Section~\ref{sec:synth_4Dflow} to create the low-resolution velocity field $\mathbf{v}_{\text{LR}}$. However, since the high-resolution field $\mathbf{v}_{\text{HR}}$ already contains noise, no additional noise was introduced in k-space. Moreover, to ensure that $\mathbf{v}_{\text{HR}}$ could serve as a reliable ground truth, a "noisy-voxel" removal procedure was applied to the mask. Specifically, voxels were identified as noisy and removed if their local velocity vector deviated by more than a 60° angular threshold from the mean of neighboring vectors within a 5-voxel radius. For inference, $\mathbf{v}_{\text{SR}}$ was obtained using the patch-cropping procedure detailed in Section~\ref{sec:patch_extr}.

%For inference, $\mathbf{v}_{\text{SR}}$ was obtained using the patch-cropping procedure detailed in Section~\ref{sec:patch_extr}.

\subsection{Network design and training}
\label{sec:net_des_and_train}

\subsubsection{Network architecture}
\label{sec:net_arch}

\begin{figure*}[t!]
  \centering
  \includegraphics[width=0.92\linewidth]{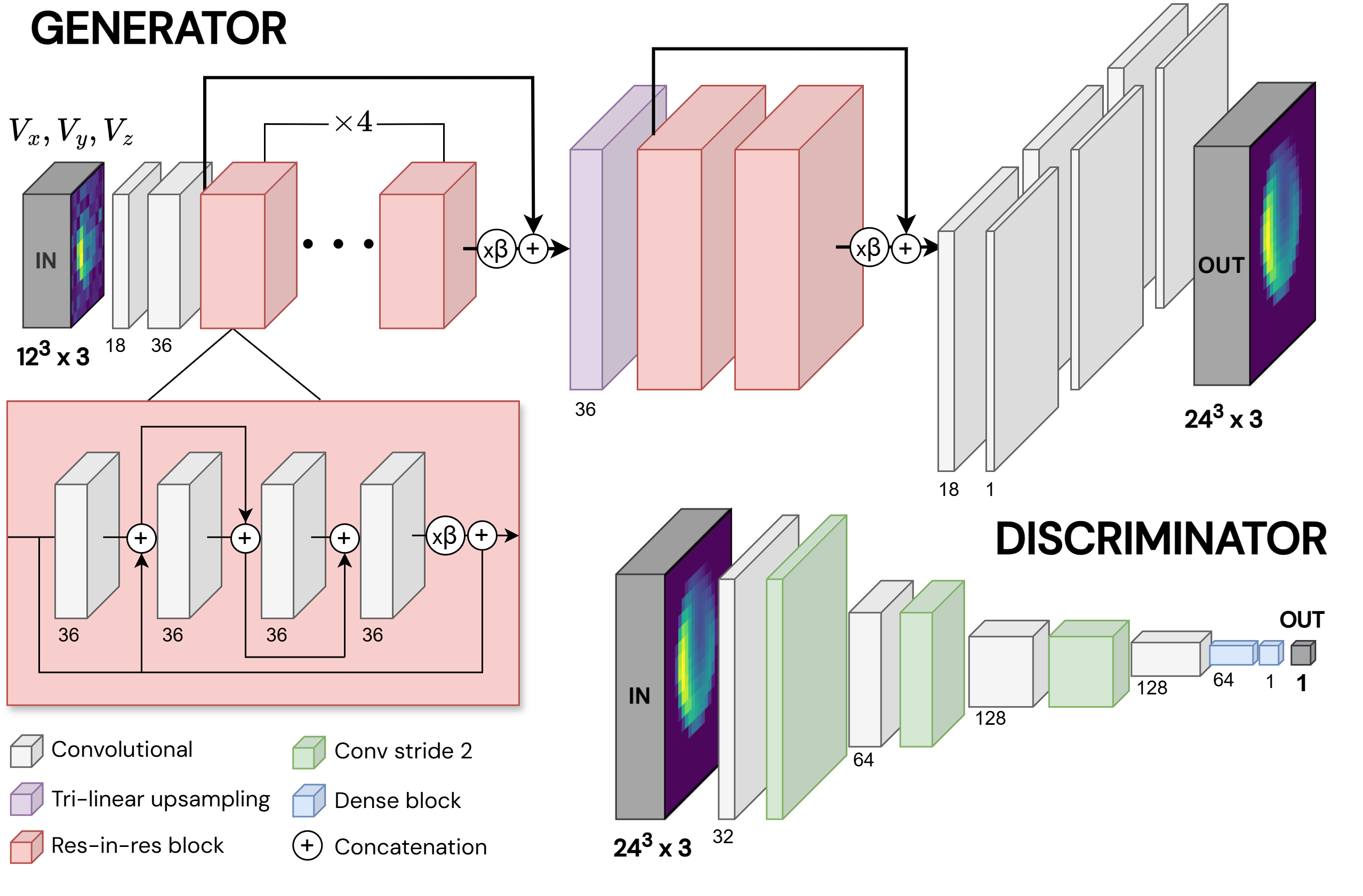}
  \caption{Overview of the proposed GAN architecture. The generator accepts cubic patches as input and incorporates residual-in-residual dense blocks, a central bicubic upsampling layer, and high-resolution blocks for predicting each velocity component. The discriminator is a CNN that progressively downsamples the input and outputs a probability distinguishing high-resolution data from generated super-resolved data.}
  \label{fig:GAN-arch}
\end{figure*}

To assess the utility of a generic GAN setup for super-resolution capabilities, we implemented a purpose-built GAN architecture with a single generator and discriminator (Figure~\ref{fig:GAN-arch}). For the generator, we adopted a design similar to 4DFlowNet~\cite{Ferdian2023Cerebrovascular}, incorporating key modifications inspired by established architectures~\cite{wang2018esrgan}. First, our generator exclusively accepts velocity data as input, in contrast to 4DFlowNet which also requires corresponding magnitude data. This design choice circumvents the challenge of simulating realistic magnitude images and notably, improved recovery performance in preliminary testing. Second, we incorporated residual-in-residual dense blocks~\cite{wang2018esrgan} to enhance gradient flow. As such, our generator \( \text{G} \) comprises four residual-in-residual dense blocks, a central upsampling layer using tri-linear interpolation, and two high-resolution blocks before branching out to predict each velocity component. 

For the discriminator, we implemented a 3D CNN to distinguish between super-resolved and high-resolution velocity fields. In brief, the network accepts a patch as input, followed by several convolutional blocks progressively downsampling the input. Finally, the extracted features are flattened and passed through fully connected layers, where the final output represents the probability of the input patch being real (\(\mathbf{x}_{\text{HR}}\)) or generated (\(\mathbf{x}_{\text{SR}} = \text{G}(\mathbf{x}_{\text{LR}})\)). 

The generator $\text{G}$ contains approximately 2.2 million trainable parameters, while the discriminator $\text{D}$ contains approximately 1.5 million.

\subsubsection{Loss functions}
\label{sec:loss_functions}

To train our GAN, we minimize a composite loss that combines data matching, adversarial, and regularization terms. The generator loss is designed to enforce accurate reconstruction of the super-resolved (SR) outputs relative to the high-resolution (HR) reference. 

In brief, it consists of three mean squared error (MSE) terms measuring discrepancies between SR and HR  in different regions: \textit{(i)} the non-fluid region, where resolved velocities are ideally zero;  \textit{(ii)} the fluid region, which is further subdivided into a near-wall area - defined as voxels adjacent to the non-fluid region (identified via binary erosion) - and the fluid core (remaining voxels). This regional decomposition is used to emphasize near-wall velocity recovery. Further, we multiply the generator adversarial loss, $L_{\mathrm{G}}$, with an adversarial loss weight $\lambda_{\mathrm{G}}$. Importantly, the adversarial weight $\lambda_{\mathrm{G}}$ balances the relative importance between the adversarial loss and the data matching terms, determining their respective contributions to the overall generator loss. An L2 regularization term, weighted by $\mu_{\mathrm{G}}$, is also included to promote parameter sparsity. 

The discriminator loss \(L_{\mathrm{Disc}}\) combines the adversarial loss \(L_{\mathrm{D}}\) with an L2 regularization term weighted by \(\mu_{\mathrm{D}}\). The overall generator and discriminator losses are expressed as: \begin{equation}
\begin{aligned}
L_{\mathrm{Gen}} &= \text{MSE}_{\mathrm{Non\text{-}Fluid}} + \text{MSE}_{\mathrm{Bound}} + \text{MSE}_{\mathrm{Core}} + \lambda_{\mathrm{G}} L_{\mathrm{G}} + \mu_{\mathrm{G}} \sum \theta_{\mathrm{G}}^2, \\
L_{\mathrm{Disc}} &= L_{\mathrm{D}} + \mu_{\mathrm{D}} \sum \theta_{\mathrm{D}}^2,
\end{aligned}
\label{eq:loss_functions}
\end{equation}

\noindent
where \(\theta_{\mathrm{G}}\) and \(\theta_{\mathrm{D}}\) denote the weights of the generator and discriminator networks, respectively. Each MSE term is computed over a distinct region \(\Omega \in \{\text{Non-Fluid}, \text{Bound}, \text{Core}\}\) according to:

\begin{equation}
\text{MSE}_{\Omega} = \frac{1}{N_\Omega} \sum_{i \in \Omega} \left\lVert \mathbf{x}^i_{\text{SR}} - \mathbf{x}^i_{\text{HR}} \right\rVert^2_2,
\label{eq:mse_region}
\end{equation}

\noindent
where \(N_\Omega\) is the number of voxels in region \(\Omega\), and \(\mathbf{x}^i_{\text{SR}}\) and \(\mathbf{x}^i_{\text{HR}}\) denote the predicted and reference velocity vectors at voxel \(i\), respectively. The regularization weights were empirically set to \(\mu_{\mathrm{G}} = 5 \cdot 10^{-7}\) and \(\mu_{\mathrm{D}} = 5 \cdot 10^{-5}\), balancing the contributions of the L2 penalties.

The performance of the defined GAN will be directly dependent on the defined loss, with training instability commonly reported as a function of the adversarial trade-offs \cite{gui2021review, saxena2021review}. In this work, we therefore choose to evaluate a variety of different adversarial loss definitions, varying  $L_{\mathrm{G}}$ and $L_{\mathrm{D}}$ in three generic ways: (1) Vanilla, (2) Relativistic, and (3) Wasserstein. These variants are evaluated in terms of their effects on training stability as well as super-resolution performance.
\vspace{8pt}

\noindent \textbf{Vanilla GAN:} The "Vanilla" formulation follows the original adversarial loss introduced by Goodfellow et al.~\cite{goodfellow2014gan}, and has been widely applied across various image synthesis tasks. In our setup, the Vanilla GAN loss can be defined as:
\begin{equation}
\begin{aligned}
L_{\mathrm{G}} &= - \mathbb{E}_{\mathbf{x}_{\text{SR}}}\left[\log \sigma( D(\mathbf{x}_{\text{SR}}))\right] \\
L_{\mathrm{D}} &= - \mathbb{E}_{\mathbf{x}_{\text{HR}}} \left[ \log \sigma(D(\mathbf{x}_{\text{HR}})) \right] - \mathbb{E}_{\mathbf{x}_{\text{SR}}} \left[ \log (1 - \sigma( D(\mathbf{x}_{\text{SR}}))) \right] ,
\end{aligned}
\label{eq:vanilla}
\end{equation}

\noindent
where $D(\cdot)$ represents the output of the discriminator before activation, and $\sigma(\cdot)$ is the sigmoid function. This formulation was successfully adopted in~\cite{ledig2017srgan}, demonstrating sharper textures compared to purely pixel-wise loss formulations. However, this standard adversarial loss is known to suffer from training instability, where training may exhibit oscillatory or divergent behavior~\cite{saxena2021review, gui2021review}. Subsequent work has therefore focused on alternative formulations to improve convergence and training dynamics. 
\vspace{8pt}

\noindent \textbf{Relativistic GAN:} The Relativistic GAN loss, introduced in \cite{jolicoeur2018relativistic}, modifies the traditional GAN framework by redefining the discriminator objective to operate in a relativistic manner. Instead of estimating the probability that a given input is real, the relativistic discriminator learns to predict the probability that a real sample is more realistic than a generated one, and vice versa. This formulation shifts the discriminator's role from \textit{absolute} classification to \textit{relative} comparison. Furthermore, the generator loss contains both real and fake samples, allowing it to benefit from the gradients of both sets. This approach was adopted in ESRGAN~\cite{wang2018esrgan}, where it was shown to improve the perceptual sharpness of super-resolved outputs in terms of edges and contrast. However, this relativistic formulation does not inherently address the aforementioned training instabilities commonly associated with adversarial learning. In our setup, we outline our relativistic loss as: \begin{equation}
\begin{aligned}
L_{\mathrm{G}} &= - \mathbb{E}_{\mathbf{x}_{\text{HR}}} \left[ \log \left( 1 - D^{\text{Rel}}(\mathbf{x}_{\text{HR}}, \mathbf{x}_{\text{SR}}) \right) \right] - \mathbb{E}_{\mathbf{x}_{\text{SR}}} \left[ \log \left( D^{\text{Rel}}(\mathbf{x}_{\text{SR}}, \mathbf{x}_{\text{HR}}) \right) \right], \\
L_{\mathrm{D}} &= - \mathbb{E}_{\mathbf{x}_{\text{HR}}} \left[\log \left( D^{\text{Rel}}(\mathbf{x}_{\text{HR}}, \mathbf{x}_{\text{SR}}) \right) \right] -\mathbb{E}_{ \mathbf{x}_{\text{SR}}} \left[ \log \left( 1 - D^{\text{Rel}}(\mathbf{x}_{\text{SR}}, \mathbf{x}_{\text{HR}}) \right) \right], 
 \end{aligned}
\label{eq:relativistic}
\end{equation}

\noindent
$\text{where} \, \, D^{\text{Rel}}(\mathbf{x}_{\text{HR}}, \mathbf{x}_{\text{SR}}) = \sigma \left( D(\mathbf{x}_{\text{HR}}) - \mathbb{E}_{\mathbf{x}_{\text{SR}}} [ D(\mathbf{x}_{\text{SR}}) ] \right).$

\vspace{8pt}

\noindent \textbf{Wasserstein GAN:} The Wasserstein GAN (WGAN)~\cite{arjovsky2017wasserstein} replaces the standard discriminator loss with the Earth Mover (Wasserstein-1) distance. This reformulation aims to improve training stability by providing smoother, more informative gradients, thus mitigating issues such as mode collapse and vanishing gradients. Unlike standard GANs, the WGAN discriminator outputs unbounded, real-valued scores rather than probabilities, allowing the loss to remain meaningful even when the generator distribution is far from the real data distribution. To ensure the theoretical validity of the Wasserstein distance, the discriminator is required to be 1-Lipschitz continuous: a constraint that promotes well-behaved gradients and mitigates erratic updates. In the original formulation~\cite{arjovsky2017wasserstein}, this condition was enforced by weight clipping, which however can restrict model capacity and lead to optimization difficulties. To overcome these limitations, Gulrajani et al.~\cite{gulrajani2017improvedwgan} proposed a gradient penalty term enforcing the Lipschitz constraint by instead penalizing the norm of the discriminator gradients on random interpolations between real and generated samples. The authors showed that this regularization yields smoother gradients for both the discriminator and generator, resulting in more stable convergence and improved training dynamics. 

In our setup, we define our Wasserstein loss as: \begin{equation}
\begin{aligned}
L_{\mathrm{G}} &= \mathbb{E}_{\mathbf{x}_{\text{SR}}} \left[ D(\mathbf{x}_{\text{SR}}) \right], \\
L_{\mathrm{D}} &=  \mathbb{E}_{\mathbf{x}_{\text{HR}}} \left[ D(\mathbf{x}_{\text{HR}}) \right] 
- \mathbb{E}_{\mathbf{x}_{\text{SR}}} \left[ D(\mathbf{x}_{\text{SR}}) \right] 
+ L_{\text{GP}},
\end{aligned}
\label{eq:wgan}
\end{equation}
$\text{where} \,\, L_{\text{GP}} = \lambda \, \mathbb{E}_{\hat{\mathbf{x}}} \left[ \left( \left\| \nabla_{\hat{\mathbf{x}}} D(\hat{\mathbf{x}}) \right\|_2 - 1 \right)^2 \right], \, \lambda = 10,\, \text{and}\,\,\hat{\mathbf{x}} = \beta \mathbf{x}_{\text{HR}} + (1 - \beta) \mathbf{x}_{\text{SR}}$, where $\beta$ is drawn from a uniform distribution $\mathcal{U}[0, 1]$.

\subsubsection{Training setup}
\label{sec:train_details}
Beyond the architectural and loss definitions mentioned above, a few additional implementation details are provided below. 
\vspace{8pt}

\noindent\textbf{Discriminator input masking:} To prevent the discriminator from trivially distinguishing real from generated images based on background content, we include a velocity segmentation mask as input prior to discriminator evaluation. Specifically, we set all voxels outside the fluid region to zero in the super-resolved images $\mathbf{x}_{\text{SR}}$ before feeding them to the discriminator. This ensures consistency with the high-resolution reference images, where the non-fluid regions are identically zero, and forces the discriminator to focus on meaningful velocity upsampling rather than background noise suppression.
\vspace{8pt}

\noindent\textbf{Two-stage training:} To enhance adversarial training stability and improve convergence, we adopt a two-stage training strategy as proposed in ESRGAN~\cite{wang2018esrgan}. In the first stage, the generator is pre-trained for 100 epochs using only the data matching loss terms, enabling the generator to learn some of the tasked super-resolution capacity before adversarial feedback is introduced. In the second stage, the generator is fine-tuned and the discriminator trained jointly for an additional 100 epochs using the full adversarial loss formulations. This warm-start approach improves discriminator stability by avoiding early overfitting to poor generator outputs, and allows it to focus on texture discrimination. Furthermore, pre-training allows for network weight interpolation, balancing CNN-based and GAN-based loss (further described below).
\vspace{8pt}

\noindent\textbf{Network weight interpolation:}
Despite the demonstrated effectiveness of adversarial training in enhancing perceptual quality and contrast \cite{ledig2017srgan, sajjadi2017enhancenet}, the potential introduction of high-frequency noise or hallucinations remains a significant concern \cite{blau2018perception, liang2022details}. To assess this trade-off in our super-resolution 4D Flow MRI setting, we chose to implement a network weight interpolation strategy inspired by ESRGAN~\cite{wang2018esrgan}. This approach allows for controlling the influence of the adversarial loss post-training, enabling an interpretable way of balancing contrast enhancement and data consistency. Specifically, we implemented an interpolation between the weights of our generator (GAN-Gen; trained using only data-matching terms), and a GAN-based generator (e.g., WGAN; obtained by fine-tuning with full adversarial loss). The weights of the resulting interpolated generator are then given by:
\begin{equation} \theta_{\mathrm{G}}^{\text{INTERP}} = (1 - \alpha) \theta_\mathrm{G}^{\text{PSNR}} + \alpha \theta_\mathrm{G}^{\text{GAN}}, \quad \alpha \in [0, 1], \end{equation}
\noindent where $\alpha$ is a scalar parameter controlling the degree of interpolation. In brief, this strategy offers several advantages. First, it allows us to separate the effects of adversarial training without re-training the model. Second, it enables smooth control over the generator’s behavior, potentially yielding intermediate networks with optimized trade-off. 

\subsection{Performance evaluation}
\label{sec:perf_eval}

\subsubsection{Network validation and comparison}
%\noindent\textbf{Validation.}
To assess super-resolution performance, we performed quantitative evaluation through voxel-wise comparisons between $\mathbf{v}_{\text{SR}}$ and $\mathbf{v}_{\text{HR}}$. For the \textit{in-silico} data, an isolated test set that was withheld from training and validation. Following synthetic data generation as per Section~\ref{sec:synth_4Dflow}, paired test data at high:low-resolution of 0.5:1.0~mm isotropic was generated, both with an effective temporal resolution of 10~ms.

Following Section~\ref{sec:loss_functions}, three different adversarial networks were trained and tested: one GAN using the default vanilla loss function (Vanilla), one GAN using a relativistic loss function (Relativistic), and one using a Wasserstein loss function (WGAN). With the generator architecture identical across the three variations, we are hence able to systematically assess the influence of adversarial loss formulation on training stability and super-resolution capability. 

In addition to this, two reference networks were also trained and included. First, to isolate the contribution of adversarial training, we included a non-adversarial baseline (GAN-Gen), where the generator is continuously trained without any adversarial influence and using data-matching loss only. This provides insight into how performance can be attributed to adversarial fine-tuning. Second, we also compare GAN performance to the reference residual 4DFlowNet \cite{Ferdian2023Cerebrovascular}; a state-of-the-art residual network in the super-resolution 4D Flow domain. To ensure consistent comparison, 4DFlowNet was here retrained from scratch using the very same training data as the rest of the networks, with the very same data and training procedure utilized. 

\subsubsection{Evaluation Metrics}
To quantify accuracy, we employed several commonly used error metrics: mean relative error (MRE), mean absolute error (MAE), vector normalized root mean square error (vNRMSE), and direction error (DE). Additionally, linear regression analysis was performed between the high-resolution reference and super-resolved velocities in each spatial direction to compute the regression slope ($k$) and the coefficient of determination ($R^2$). The error metrics are defined as below: \begin{equation} \begin{aligned} \text{MRE} &= \frac{1}{N} \sum_{i=1}^{N} \tanh\left(\dfrac{\left\| \mathbf{v}_{\text{SR}}^i - \mathbf{v}_{\text{HR}}^i \right\|_2}{\left\| \mathbf{v}_{\text{HR}}^i \right\|_2 + \epsilon}\right) \\ 
\text{MAE} &= \frac{1}{N} \sum_{i=1}^{N} \left\| \mathbf{v}_{\text{SR}}^i - \mathbf{v}_{\text{HR}}^i \right\|_2 \\ 
\text{vNRMSE} &= \frac{1}{\max_i{\left\| \mathbf{v}_{\text{HR}} \right\|_2}} \sqrt{\frac{1}{N}\sum_{i=1}^{N} \left\| \mathbf{v}_{\text{SR}}^i - \mathbf{v}_{\text{HR}}^i \right\|^2_2}  \\ 
\text{DE} &= \frac{1}{N} \sum_i^N\left(1 - \frac{\left| \mathbf{v}_{\text{SR}}^i \cdot \mathbf{v}_{\text{HR}}^i \right|}{\left\| \mathbf{v}_{\text{SR}}^i \right\|_2 \cdot \left\| \mathbf{v}_{\text{HR}}^i \right\|_2 + \epsilon} \right) \end{aligned} \label{eq:error_metrics} \end{equation} \noindent where $\mathbf{v}_{\text{SR}}^i$ and $\mathbf{v}_{\text{HR}}^i$ denote the super-resolved and high-resolution velocity vectors at voxel $i$, respectively, $\epsilon=10^{-6}$, and $N$ is the total number of fluid-containing voxels.

All metrics are computed across the full cardiac cycle as well as at peak systolic flow. Furthermore, we isolate the boundary (identified using binary erosion) and core (remaining voxels) regions to separately assess performance. For the \textit{in-vivo} datasets, we report metrics only in the full fluid region due to uncertain boundary delineation and disproportionately low SNR at the wall. Further, as discrete synthetic time frames were randomly assigned either low or high TSNR (see Section~\ref{sec:synth_4Dflow}), \textit{in-silico} evaluations were also separated into high SNR time frames (TSNR=10--12) and low SNR time frames (TSNR=2--4), allowing for SNR-specific quantification.

\subsubsection{Feature distribution analysis}
\label{sec:feature_distr}
In addition to the above scalar metrics, to qualitatively assess how adversarial training influences internal feature representations, we also performed a network feature distribution analysis using principal component analysis (PCA). Analyzing internal representations can reveal differences in how networks encode information in latent space, and has been commonly used to interpret the effectiveness of the training process \cite{islam2023revealing}. Feature activations were extracted from two locations in the generator network: (1) at the output of the final low-resolution residual-in-residual block prior to tri-linear interpolation (“middle”), and (2) at the output of the final high-resolution block before branching into the three velocity components (“end”). For each location, features were collected from 10,000 randomly sampled training patches. The resulting high-dimensional feature vectors were projected onto the first two principal components to enable visual comparison across networks.

\label{sec:training_stability}
\begin{figure*}[!t]
  \centering
  \includegraphics[width=0.95\linewidth]{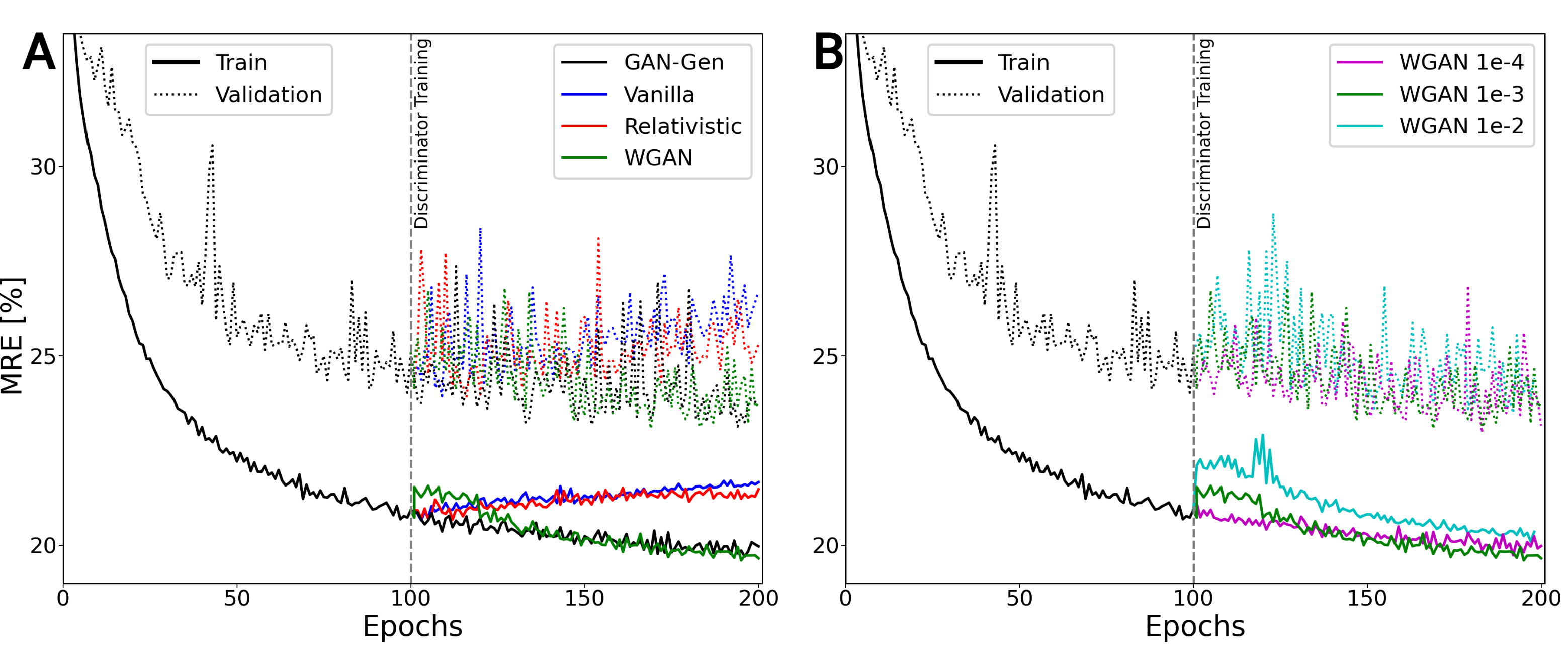}
  \caption{\textbf{(A)} Training and validation mean relative error (MRE) over 200 epochs. The first 100 epochs correspond to generator-only training (GAN-Gen). For subsequent epochs, adversarial losses are introduced. \textbf{(B)} Effect of varying adversarial loss weight ($\lambda_{\text{G}}$) on WGAN loss curves.}
  \label{fig:Training_curves_1}
\end{figure*}

\subsection{Computational implementation}
All networks were implemented in TensorFlow (v2.10). We used the Adam optimizer with a learning rate of $1 \times 10^{-4}$ and a batch size of 20 across all experiments. Training was performed on an NVIDIA A100 Tensor Core GPU with 40 GB of memory. The total training time was approximately 24 hours for 100 epochs of generator training, followed by an additional 25 hours of training for each GAN variant using the full adversarial loss formulation.
%\enlargethispage{\baselineskip}

\section{Results}

\subsection{Training stability analysis}

Figure~\ref{fig:Training_curves_1}A displays training and validation mean relative error (MRE) behavior across 200 epochs. As observed, both Vanilla and Relativistic GANs exhibit unstable training behavior, with MRE diverging after the adversarial loss is activated. In contrast, the WGAN shows a momentary increase in error followed by stabilization to the GAN-Gen baseline, indicating a more robust optimization behavior. While the MRE for the validation set remains consistently higher than the training MRE across all networks, performance follows the same overall trends. Figure~\ref{fig:Training_curves_1}B shows the same MRE training curves, however when varying the generator adversarial loss of the WGAN setup for weight $\lambda_{\text{G}} \in \{10^{-4}, 10^{-3}, 10^{-2}\}$. Among these, $\lambda_{\text{G}} = 10^{-3}$ provided optimal training performance, with an MRE of 19.6\% after 200 epochs. This setting demonstrated a clear influence of adversarial learning while preserving competitive validation performance.

\begin{figure*}[t!]
  \centering
\includegraphics[width=0.94\linewidth]{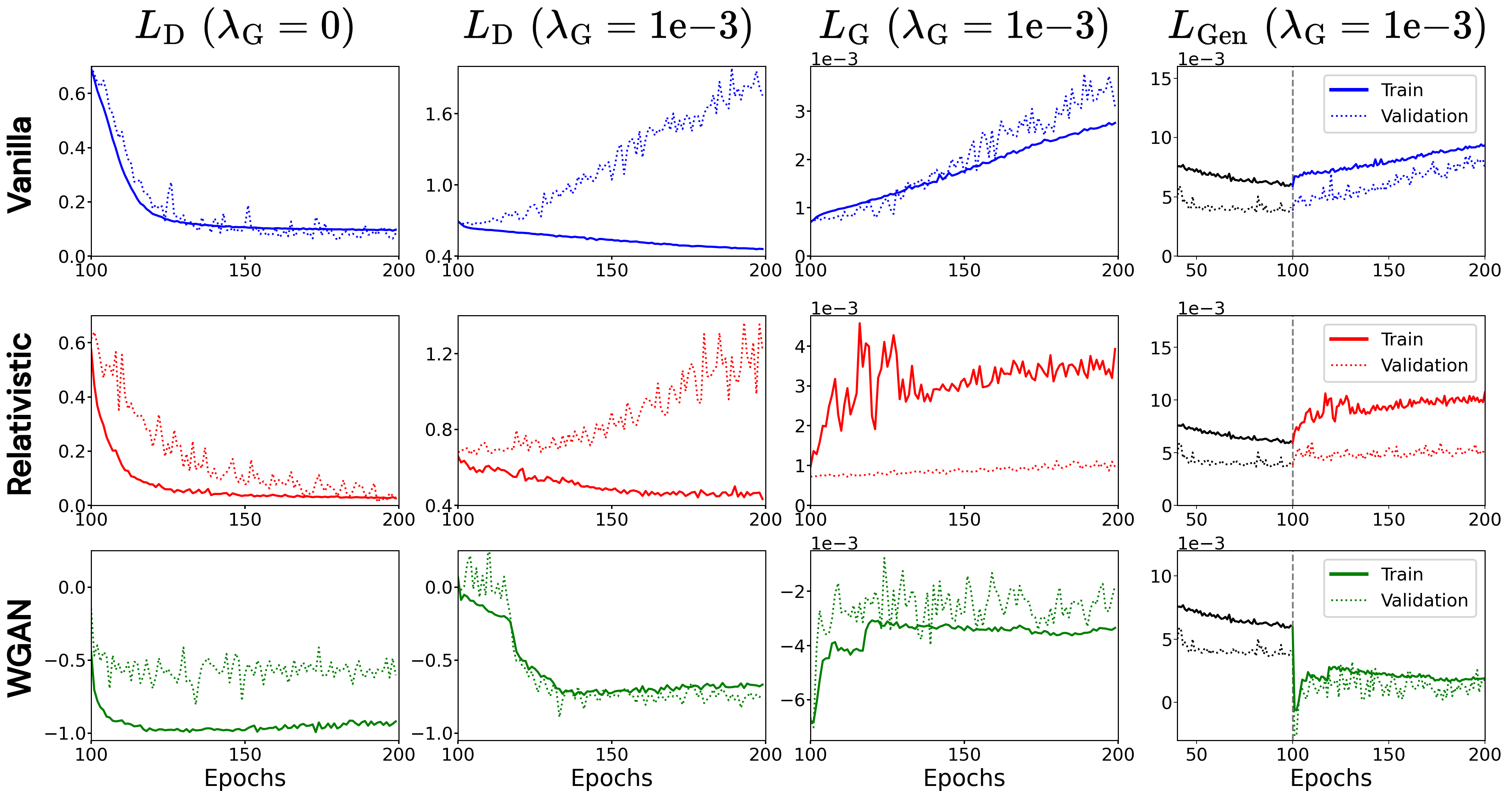} \caption{Discriminator ($L_{\text{D}}$) and generator loss ($L_{\text{G}}$ and $L_{\text{Gen}}$) trajectories for Vanilla, Relativistic, and WGANs under two regimes: without adversarial feedback ($\lambda_{\text{G}} = 0$, first column), and with adversarial feedback ($\lambda_{\text{G}} = 10^{-3}$, remaining columns).}
  \label{fig:Training_curves2}
\end{figure*}

Continuing into detailed dissemination of training behavior, Figure~\ref{fig:Training_curves2} shows the evolution of discriminator and generator losses for Vanilla, Relativistic, and  WGANs across two forms of training: one exclusively using the discriminator loss ($L_{\text{D}}$) without \textit{any} feedback to the generator ($\lambda_{\text{G}}=0$); and one with adversarial feedback activated ($\lambda_{\text{G}}=10^{-3}$). As observed, in the discriminator-only setup, all networks exhibit similar converging behavior with Vanilla and Relativistic GANs both approaching zero-loss, and WGAN approaching a corresponding negative convergence value (reflecting the Wasserstein distance formulation) (Figure~\ref{fig:Training_curves2}, first column).  
On the contrary, when adversarial feedback is enabled, both Vanilla and Relativistic GANs exhibit typical divergent behavior with validation losses increasing steadily with progressing epochs. The same does, however, not hold true for the WGAN setup where instead training and validation losses both decrease rapidly at earlier epochs before jointly plateauing at stabilized convergence in the later stages of training (Figure~\ref{fig:Training_curves2}, second column).
Supporting this, the generator adversarial loss $L_{\text{G}}$ of the same setup ($\lambda_{\text{G}} = 10^{-3}$) highlights the very same behavior where errors for Vanilla and Relativistic GAN setups grow erratically (indicating the generator's ability to deceive the discriminator), whereas corresponding WGAN curves stabilize after an initial increase, consistent with effective generator-discriminator interaction (Figure~\ref{fig:Training_curves2}, third column).     

Finally, the fourth column of Figure~\ref{fig:Training_curves2} presents the total generator loss $L_{\text{Gen}}$, combining $L_{\text{G}}$ with the baseline data matching loss, again highlighting divergence patterns for Vanilla and Relativistic GANs, while WGAN maintains a stable trajectory after the initial transient stabilization phase. 

\begin{figure*}[!b]%[ht]
  \centering
  \includegraphics[width=0.90\linewidth]{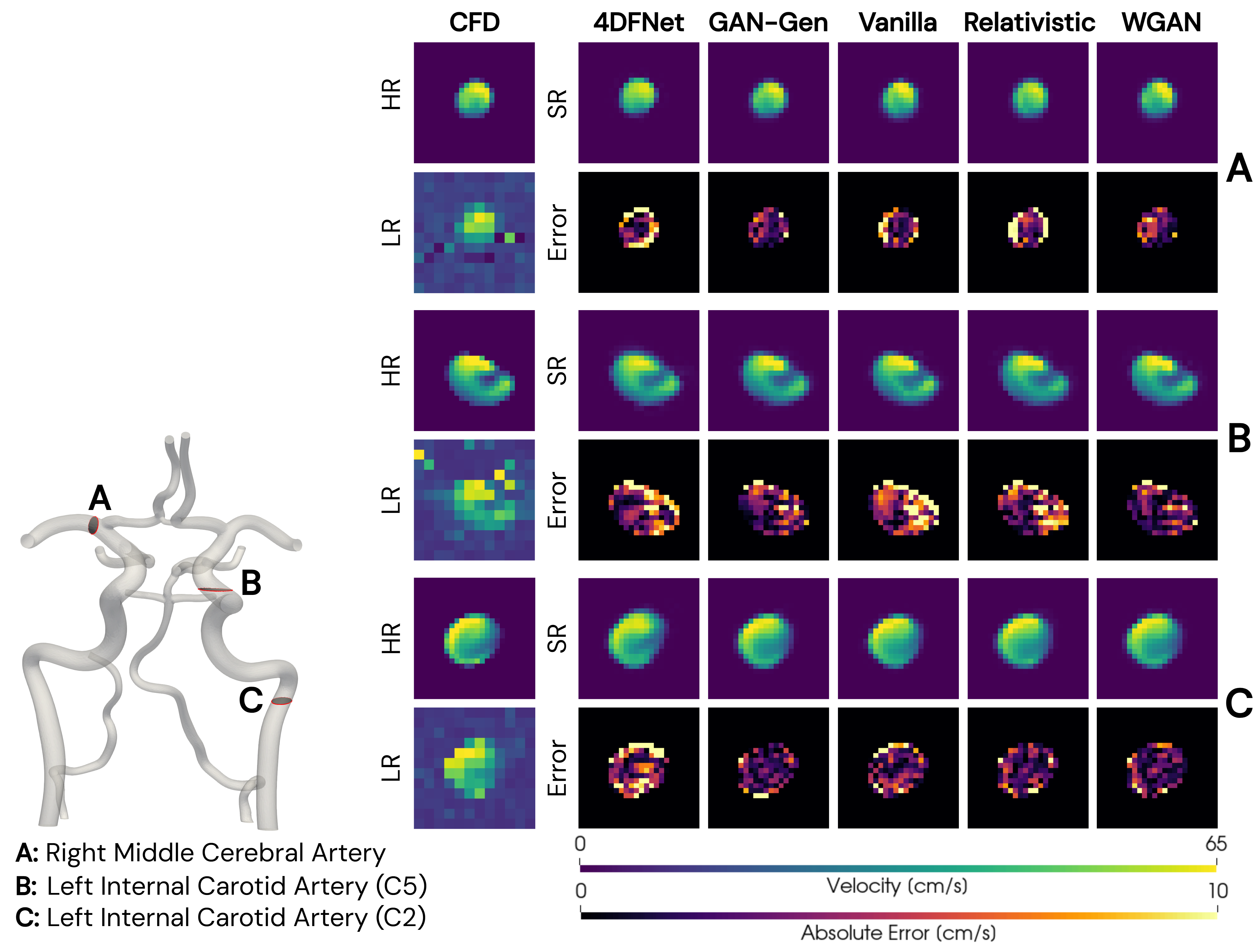}
  \caption{Qualitative comparison between low resolution (LR), high resolution (HR) and super-resolution (SR) at three anatomical locations during peak systolic flow, along with corresponding absolute error maps (HR vs SR).} 
  \label{fig:Patch_vis}
\end{figure*}

\begin{table*}[!t]
\centering
\small
\caption{Average performance metrics for all networks across core and boundary regions, computed over the full cardiac cycle. Best values are highlighted in bold.}
\label{tab:multirow}
\begin{tabular}{l l c c c c c c}
\toprule
\textbf{Region} & \textbf{Model} & $\, \,$ MRE [\%] & MAE & vNRMSE
& $\, \,$ DE [\%] & \(k\) & \(R^2\) \\
\midrule
\multirow{5}{*}{\textbf{Boundary}} 
 & 4DFlowNet
   & 39.73  & 0.0528 & 0.0962 & \textbf{4.05}  
   & (0.858, 0.870, 0.896)
   & (0.764, 0.791, 0.812) \\
 & GAN-Gen
   & 38.93  & 0.0443 & 0.0719 & 6.72
   & (0.883, 0.887, 0.894)
   & (0.878, 0.874, 0.866) \\
 & Vanilla
   & 43.45  & 0.0513 & 0.0806 & 7.95
   & (0.845, 0.847, 0.791)
   & (0.853, 0.856, 0.829) \\
 & Relativistic
   & 41.51  & 0.0487 & 0.0780 & 6.42
   & (\textbf{0.906}, 0.919, 0.890)
   & (0.856, 0.866, 0.841) \\
 & WGAN
   & \textbf{38.30}  & \textbf{0.0428} & \textbf{0.0690} & 5.85
   & (0.897, \textbf{0.935}, \textbf{0.900})
   & (\textbf{0.890}, \textbf{0.887}, \textbf{0.876}) \\
\midrule
\multirow{5}{*}{\textbf{Core}}
 & 4DFlowNet
   & 12.15  & 0.0366 & 0.0655 & \textbf{0.69}
   & (0.956, 0.930, 0.952)
   & (0.963, 0.969, 0.965) \\
 & GAN-Gen
   & 11.77  & \textbf{0.0319}  & \textbf{0.0501}  & 1.48
   & (\textbf{0.984}, 0.966, \textbf{0.976})
   & (\textbf{0.981}, 0.980, 0.973) \\
 & Vanilla
   & 15.43  & 0.0416 & 0.0635 & 1.34
   & (0.972, 0.957, 0.942)
   & (0.972, 0.973, 0.960) \\
 & Relativistic
   & 12.77  & 0.0356 & 0.0551 & 1.56
   & (0.979, 0.958, 0.970)
   & (0.975, 0.978, 0.970) \\
 & WGAN
   & \textbf{11.65}  & 0.0322 & 0.0503 & 1.38
   & (0.964, \textbf{0.972}, 0.963)
   & (\textbf{0.981}, \textbf{0.983}, \textbf{0.975}) \\
\bottomrule
\end{tabular}
\label{table:overall_metrics}
\end{table*}

\subsection{Qualitative and Quantitative super-resolution performance}

\subsubsection{Overall performance}

Qualitative comparison between the trained networks is highlighted in Figure~\ref{fig:Patch_vis}, illustrating representative cross-sections from patches extracted across three vascular sections during peak systole. 
As visually apparent, all SR outputs closely resemble the HR reference with background noise effectively removed, demonstrating successful overall super-resolution performance. However, distinct differences can be observed in accuracy and boundary recovery in the corresponding error maps. While 4DFlowNet consistently exhibits higher errors (with the exception of DE), particularly close to the vessel boundaries, these are reduced when shifting to the GAN-Gen setup. In contrast, both Vanilla and Relativistic GANs introduce higher errors, again observed most clearly at the vessel boundary. As congruent with Section~\ref{sec:training_stability}, WGAN however mitigates this trend with overall output comparable to the GAN-Gen setup. Moving into quantitative descriptions, Table~\ref{table:overall_metrics} provides error metrics across all networks, supporting the visual trends observed. In the core region, GAN-Gen and WGAN achieve near-equivalent performance with an MRE of 11.77\% vs. 11.65\%, and a vNRMSE of 0.0501 vs. 0.0503, whereas the reference 4DFlowNet along with the Vanilla and Relativistic GAN setups all show comparably worse performance (MRE: 12.15, 15.43 and 12.77 \%; vNRMSE: 0.0655, 0.0635, 0.0551). In the boundary region, WGAN yields slightly lower boundary errors as compared to GAN-Gen with an MRE of 38.30\% vs. 38.93\% and vNRMSE of 0.0690 vs 0.0719. Again, alternative 4DFlowNet, Vanilla, and Relativistic GAN setups are less accurate. 

\subsubsection{SNR-specific performance}

\begin{figure*}[b!]
  \centering
  \includegraphics[width=0.75\linewidth]{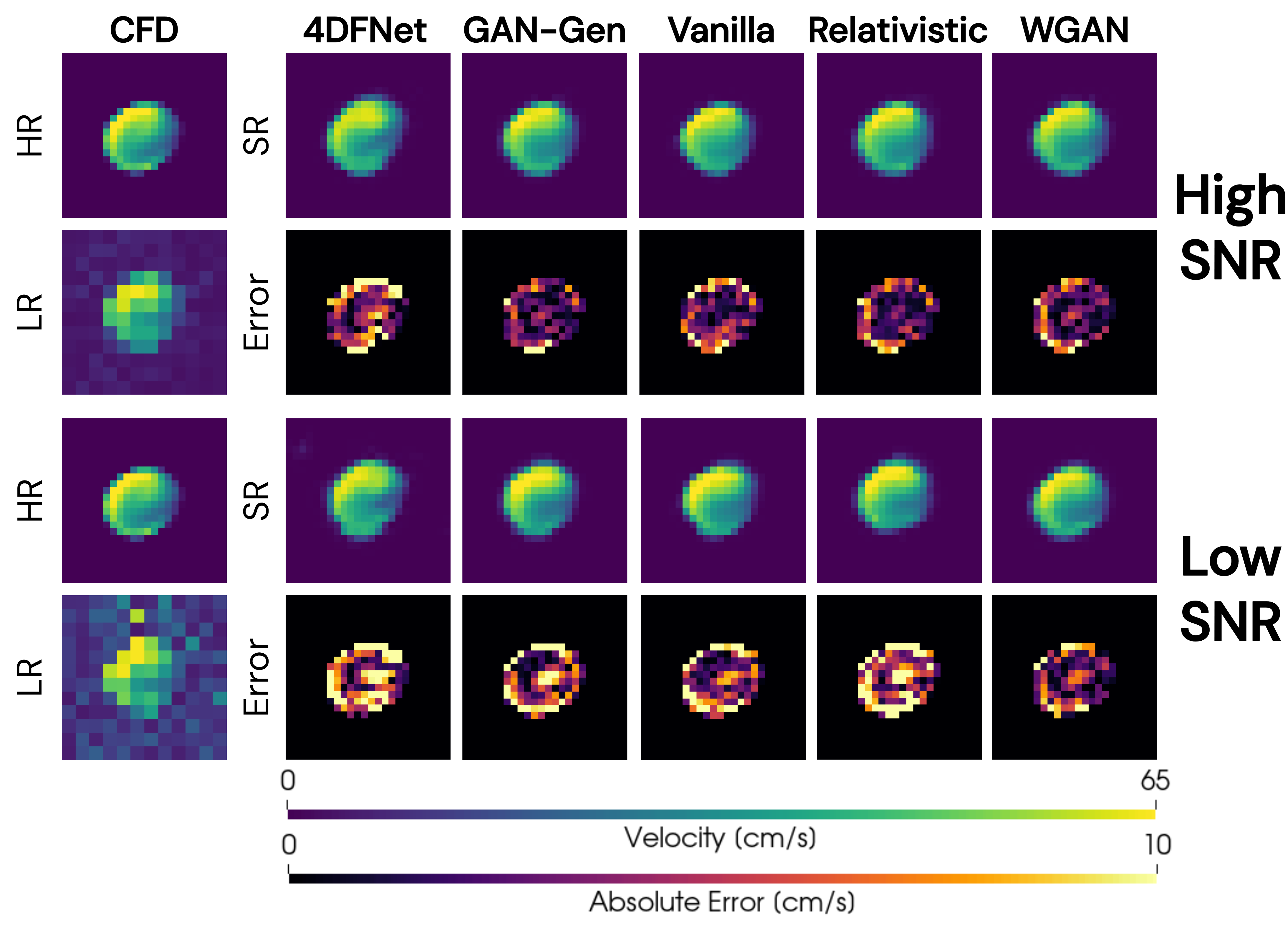}
  \caption{Qualitative comparison between low resolution (LR), high resolution (HR) and super-resolution (SR) with corresponding absolute error maps (HR vs SR) for a cross-sectional slice during peak systolic flow, under high and low SNR conditions.}
  \label{fig:High_Low_SNR_vis}
\end{figure*}

\begin{figure*}[b!]
  \centering
  \includegraphics[width=0.95\linewidth]{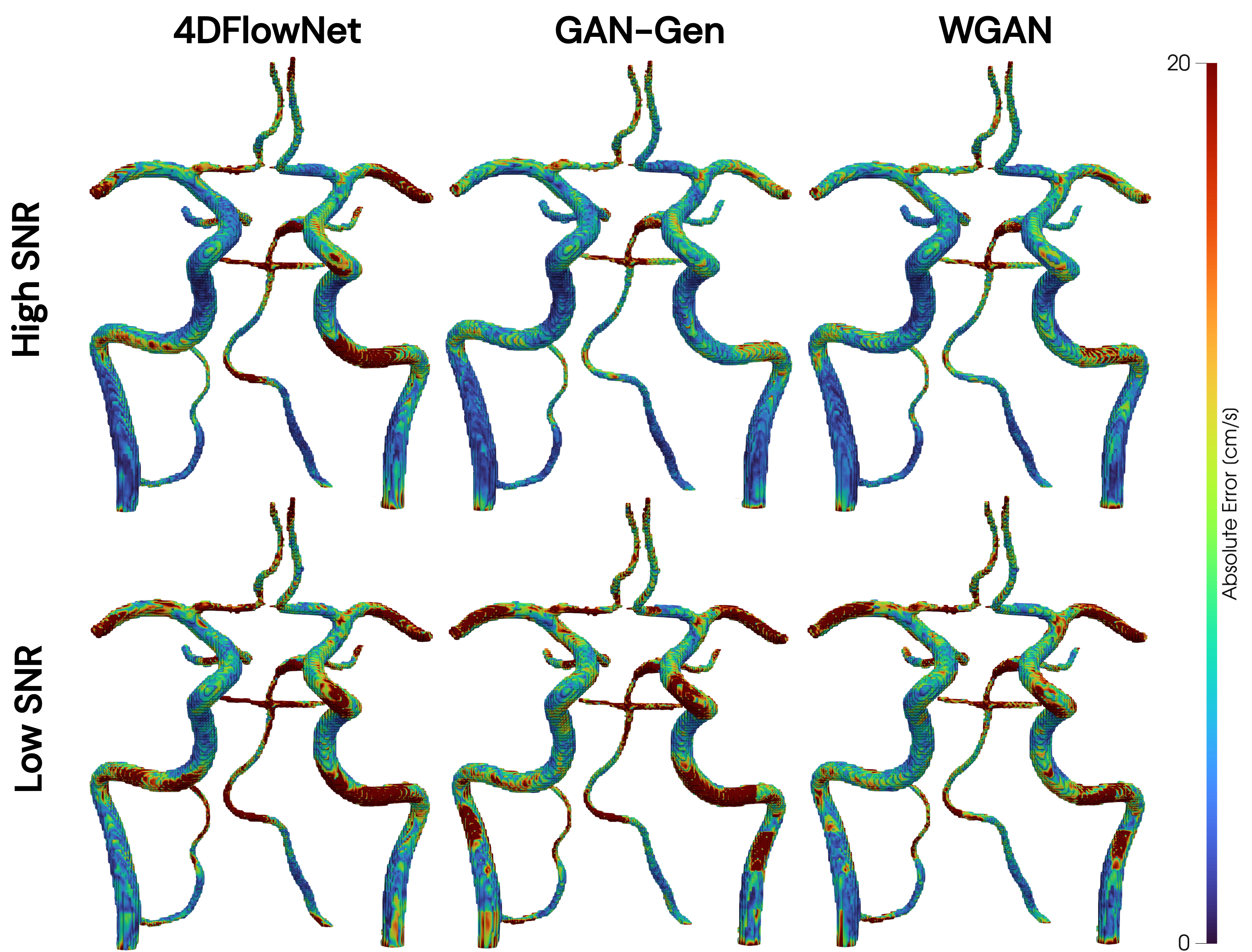}
  \caption{Boundary error visualization for 4DFlowNet, GAN-Gen, and WGAN under high and low SNR. Error maps were extracted during peak systole and visualized using ParaView~\cite{ahrens2005paraview}.}
  \label{fig:Full_Error_vis}
\end{figure*}

Figure~\ref{fig:High_Low_SNR_vis} presents qualitative reconstructions and corresponding error maps from a representative cross-sectional slice under high and low SNR conditions at peak systole, allowing for comparison of model behavior across different noise levels. Notably, at high SNR all networks recover velocity representations closely matching the high-resolution reference data; however where 4DFlowNet exhibits elevated boundary errors, these are visually suppressed across all GAN-setups. At low SNR, all networks exhibit visibly increased reconstruction errors, particularly near boundaries. However, here WGAN indicates the lowest errors near the boundary. Complementing this, Tables~\ref{Table:highSNR_combined}-\ref{Table:lowSNR_combined} present quantitative error metrics for the same assessment. As noted visually, at high SNR (Table~\ref{Table:highSNR_combined}) GAN-Gen consistently achieves the highest accuracy in both core and boundary regions, with 26 out of 40 error metrics (including e.g. MRE, MAE, vNRMSE) optimally recovered using the GAN-Gen setup. Conversely, 4DFlowNet indicates the poorest performance, with 18 out of 40 metrics indicating the highest errors at high SNR. Continuing into low SNR conditions (Table~\ref{Table:highSNR_combined}), a shift in performance is observed, with WGAN now consistently outperforming other networks with 33 out of 40 error metrics optimally recovered, including both boundary and core regions across peak systole and the average cardiac cycle. In comparison, both GAN-Gen and 4DFlowNet exhibit elevated boundary errors and higher variance across directional and regression-based metrics, reflecting poorer generalization under noise. 

To further illustrate boundary-specific performance, Figure~\ref{fig:Full_Error_vis} visualizes absolute boundary errors along an entire cerebrovascular model, stratified as a function of SNR. As shown, at high SNR 4DFlowNet exhibits the highest boundary errors, exceeding 20 cm/s across various vascular sections. In comparison, both GAN-Gen and WGAN show notable improvements along vascular edges. Errors then increase in both magnitude and distribution at low SNR; however here WGAN indicates robustness in maintaining lower errors across a majority of the model.

\begin{table*}[h!]
\centering
\footnotesize
\caption{Performance metrics for all networks across core and boundary regions using only high SNR timesteps. Computed both over the full cardiac cycle (top) and during peak systole (bottom).}
\label{Table:highSNR_combined}
\begin{tabular}{l l c c c c c c}
\toprule
\textbf{Region} & \textbf{Model} & $\, \,$ MRE [\%] & MAE & vNRMSE
& $\, \,$ DE [\%] & \( k \) & \(R^2\) \\
\midrule
\multirow{5}{*}{\textbf{Boundary}} 
 & 4DFlowNet
   & 37.14 & 0.0467 & 0.0910 & \textbf{3.49}
   & (0.877, 0.884, 0.910) & (0.785, 0.812, 0.843) \\
 & GAN-Gen
   & \textbf{34.98} & \textbf{0.0351} & 0.0591 & 5.70
   & (0.894, 0.895, 0.906) & (\textbf{0.927}, 0.912, \textbf{0.914}) \\
 & Vanilla
   & 39.74 & 0.0416 & 0.0676 & 6.55
   & (0.883, 0.873, 0.817) & (0.907, 0.896, 0.890) \\
 & Relativistic
   & 37.79  & 0.0392 & 0.0651 & 5.05
   & (\textbf{0.925}, 0.935, 0.904)  & (0.908, 0.902, 0.893) \\
 & WGAN
   & 35.16  & 0.0354 & \textbf{0.0588} & 4.95
   & (0.912, \textbf{0.945}, \textbf{0.914}) & (\textbf{0.927}, \textbf{0.916}, 0.912) \\
\midrule
\multirow{5}{*}{\textbf{Core}}
 & 4DFlowNet
   & 10.75 & 0.0322 & 0.0610 & \textbf{0.56}   
   & (0.962, 0.935, 0.958) & (0.969, 0.973, 0.971) \\
 & GAN-Gen
   & \textbf{9.58} & \textbf{0.0252} & \textbf{0.0402}  & 1.13
   & (\textbf{0.995}, 0.976, \textbf{0.985}) & (\textbf{0.989}, \textbf{0.988}, \textbf{0.984}) \\
 & Vanilla
   & 13.18  & 0.0343 & 0.0530 & 0.99 
   & (0.989, 0.968, 0.953) & (0.983, 0.982, 0.977) \\
 & Relativistic
   & 10.71 & 0.0291 & 0.0457 & 1.22 
   & (0.992, 0.968, 0.976) & (0.985, 0.985, 0.980) \\
 & WGAN
   & 10.00  & 0.0271 & 0.0431 & 1.11
   & (0.974, \textbf{0.980}, 0.970) & (0.988, \textbf{0.988}, 0.982) \\
\midrule \midrule
\textbf{Region} & \textbf{Model} & $\, \,$ MRE [\%] & MAE & vNRMSE
& $\, \,$ DE [\%] & \( k \) & \(R^2\) \\
\midrule
\multirow{5}{*}{\textbf{Boundary}} 
 & 4DFlowNet
   & 38.53  & 0.0980 & 0.1179 & 3.98
   & (0.787, 0.885, 0.879) & (0.759, 0.811, 0.823) \\
 & GAN-Gen
   & \textbf{32.61} & \textbf{0.0697} & \textbf{0.0795} & 3.21
   & (0.881, 0.876, 0.901) & (\textbf{0.914}, 0.892, \textbf{0.907}) \\
 & Vanilla
   & 37.25 & 0.0822 & 0.0892 & 5.34
   & (0.865, 0.889, 0.842) & (0.886, 0.877, 0.880) \\
 & Relativistic
   & 34.58 & 0.0770 & 0.0866 & \textbf{2.91}
   & (0.864, 0.883, 0.845) & (0.891, 0.886, 0.882) \\
 & WGAN
   & 32.69 & 0.0700 & 0.0807 & 2.96
   & (\textbf{0.887}, \textbf{0.930}, \textbf{0.912}) & (0.905, \textbf{0.895}, 0.905) \\
\midrule
\multirow{5}{*}{\textbf{Core}}
 & 4DFlowNet
   & 12.14 & 0.0643 & 0.0751 & 0.610
   & (0.958, 0.932, 0.928) & (0.958, 0.969, 0.959) \\
 & GAN-Gen
   & \textbf{9.35} & \textbf{0.0469} & \textbf{0.0496} & \textbf{0.402}
   & (\textbf{0.987}, 0.967, \textbf{0.976}) & (\textbf{0.985}, \textbf{0.984}, \textbf{0.980}) \\
 & Vanilla
   & 12.00 & 0.0603 & 0.0595 & 0.536
   & (0.981, \textbf{0.985}, 0.952) & (0.978, 0.980, 0.974) \\
 & Relativistic
   & 10.83 & 0.0553 & 0.0579 & 0.484
   & (0.978, 0.950, 0.958) & (0.979, 0.981, 0.972) \\
 & WGAN
   & 9.92 & 0.0506 & 0.0534 & 0.411
   & (0.975, 0.975, 0.963) & (0.982, \textbf{0.984}, 0.976) \\
\bottomrule
\end{tabular}
\end{table*}

\begin{table*}[ht!]
\centering
\footnotesize
\caption{Performance metrics for all networks across core and boundary regions using only low SNR timesteps. Computed both over the full cardiac cycle (top) and during peak systole (bottom).}
\label{Table:lowSNR_combined}
\begin{tabular}{l l c c c c c c}
\toprule
\textbf{Region} & \textbf{Model} & $\, \,$ MRE [\%] & MAE & vNRMSE
& $\, \,$ DE [\%] & \(k\) & \(R^2\) \\
\midrule
\multirow{5}{*}{\textbf{Boundary}} 
 & 4DFlowNet
   & 43.83  & 0.0645 & 0.1065 & \textbf{5.03}
   & (0.823, 0.847, 0.871)
   & (0.731, 0.757, 0.761) \\
 & GAN-Gen
   & 45.27  & 0.0619 & 0.0956 & 8.08   
   & (0.856, 0.867, 0.865) & (0.789, 0.807, 0.778) \\
 & Vanilla
   & 49.27  & 0.0698 & 0.1046 & 10.11
   & (0.777, 0.806, 0.747) & (0.753, 0.783, 0.719) \\
 & Relativistic
   & 47.21  & 0.0664 & 0.1010 & 8.24
   & (0.865, 0.888, 0.857) & (0.767, 0.804, 0.748) \\
 & WGAN
   & \textbf{43.25} & \textbf{0.0566} & \textbf{0.0875} & 7.12
   & (\textbf{0.872}, \textbf{0.916}, \textbf{0.874}) & (\textbf{0.826}, \textbf{0.837}, \textbf{0.813}) \\
\midrule
\multirow{5}{*}{\textbf{Core}}
 & 4DFlowNet
   & 14.43  & 0.0448 & 0.0737 & \textbf{0.93}
   & (0.945, 0.920, 0.942) & (0.954, 0.962, 0.953) \\
 & GAN-Gen
   & 15.44 & 0.0446 & 0.0683 & 1.90 
   & (\textbf{0.966}, 0.950, \textbf{0.959}) & (0.964, 0.967, 0.950) \\
 & Vanilla
   & 19.09 & 0.0556 & 0.0827 & 1.86
   & (0.943, 0.941, 0.922) & (0.952, 0.955, 0.927) \\
 & Relativistic
   & 16.14 & 0.0478 & 0.0718 & 1.98
   & (0.955, 0.941, 0.957) & (0.958, 0.965, 0.950) \\
 & WGAN
   & \textbf{14.40}  & \textbf{0.0418}
   & \textbf{0.0632} & 1.73
   & (0.949, \textbf{0.959}, 0.949) & (\textbf{0.971}, \textbf{0.972}, \textbf{0.960}) \\
\midrule \midrule
\textbf{Region} & \textbf{Model} & $\, \,$ MRE [\%] & MAE & vNRMSE
& $\, \,$ DE [\%] & \( k \) & \(R^2\) \\
\midrule
\multirow{5}{*}{\textbf{Boundary}} 
 & 4DFlowNet
   & 45.69 & 0.1252 & 0.1406 & 5.59
   & (0.694, 0.848, 0.817) & (0.648, 0.756, 0.720)  \\
 & GAN-Gen
   & 46.79 & 0.1245 & 0.1348 & 5.72
   & (0.771, 0.844, 0.827) & (0.708, 0.769, 0.709)  \\
 & Vanilla
   & 51.26 & 0.1408 & 0.1478 & 9.37
   & (0.721, 0.827, 0.770) & (0.654, 0.736, 0.636)  \\
 & Relativistic
   & 48.20 & 0.1320 & 0.1409 & 5.87
   & (0.758, 0.832, 0.776) & (0.693, 0.756, 0.655)  \\
 & WGAN
   & \textbf{43.72} & \textbf{0.1137} & \textbf{0.1261} & \textbf{4.66} 
   & (\textbf{0.792}, \textbf{0.889}, \textbf{0.832}) & (\textbf{0.740}, \textbf{0.785}, \textbf{0.763})\\
\midrule
\multirow{5}{*}{\textbf{Core}}
 & 4DFlowNet
   & 15.04 & 0.0800 & 0.0909 & 0.86
   & (0.917, 0.912, 0.902) & (0.933, 0.960, 0.939)  \\
 & GAN-Gen
   & 15.99 & 0.0815 & 0.0883 & 0.86
   & (\textbf{0.937}, 0.939, \textbf{0.930}) & (0.939, 0.961, 0.936)  \\
 & Vanilla
   & 19.62 & 0.1007 & 0.1050 & 1.31
   & (0.907, \textbf{0.949}, 0.893) & (0.921, 0.951, 0.899)  \\
 & Relativistic
   & 16.72 & 0.0869 & 0.0931 & 0.92
   & (0.920, 0.919, 0.920) & (0.932, 0.957, 0.933)  \\
 & WGAN
   & \textbf{14.42} & \textbf{0.0744} & \textbf{0.0808}
   & \textbf{0.75} & (0.920, 0.944, 0.922) & (\textbf{0.950}, \textbf{0.965}, \textbf{0.951})
   \\
\bottomrule
\end{tabular}
\end{table*}

\subsection{Network interpolation experiments}

\begin{figure}[ht]
  \centering
  \includegraphics[width=0.5\linewidth]{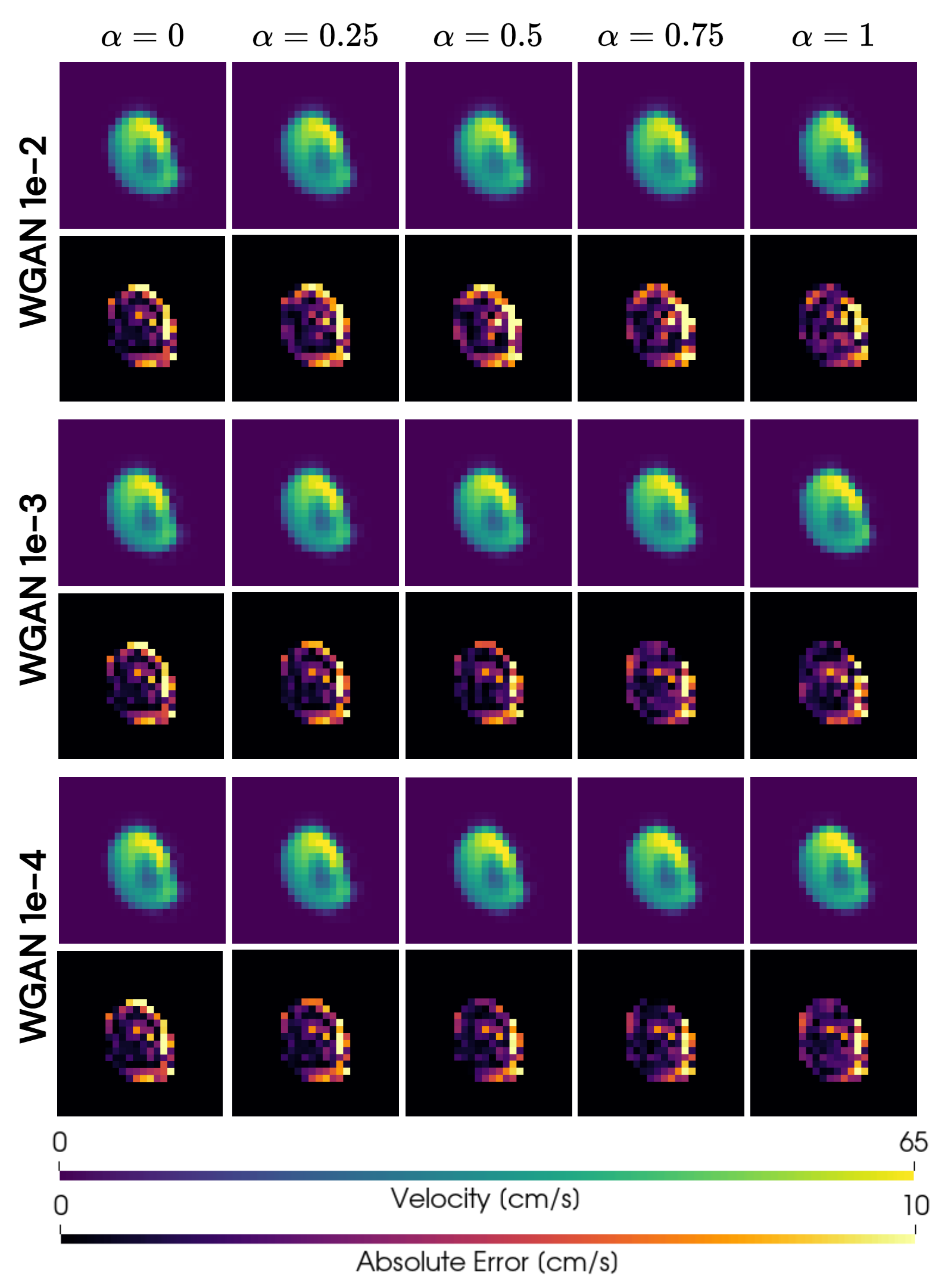}
  \caption{Representative cross-sectional velocity fields and corresponding error maps illustrating weight interpolation (\(\alpha \in [0,1]\)) between the purely CNN-based (GAN-Gen) and adversarial WGAN networks for three adversarial weight settings \(\lambda_{\text{G}} \in \{10^{-2}, 10^{-3}, 10^{-4}\}\).}
  \label{fig:interpolation_vis}
\end{figure}

To examine the effect of transitioning from a voxel-wise to a perceptual loss description, Figure \ref{fig:interpolation_vis} shows step-wise interpolation from a pure GAN-Gen setup ($\alpha=0$, representing the best-performing generator before adversarial training) to a  continually trained WGAN ($\alpha=1$). As presented, SR output appears visually similar between interpolations, with only subtle boundary improvements emerging at higher perceptual loss weighting. This is also confirmed in Table~\ref{tab:interpolation_metrics} with optimal performance observed at $\alpha = 0.75$ or $1$, although improvements are comparably modest. 

\begin{table}[ht!]
\centering
   \scriptsize
   \caption{Quantitative performance metrics (averaged over the full cardiac cycle) corresponding to the interpolated networks shown in Figure~\ref{fig:interpolation_vis}.}
        \label{tab:interpolation_metrics}
        \begin{tabular}{llccccc}
        \toprule
         & & \multicolumn{5}{c}{$\alpha$} \\
        \cmidrule(lr){3-7}
        \textbf{Metric} & \textbf{Region} & 0 & 0.25 & 0.5 & 0.75 & 1 \\
        \midrule
        \multicolumn{7}{c}{\textbf{WGAN1e-2}} \\
        \midrule
        MRE [\%] & Bound & \textbf{39.55} & 39.98 & 40.91 & 40.82 & 39.94 \\
                 & Core    & 12.33 & 12.43 & 12.87 & 12.55 & \textbf{11.55} \\
        \addlinespace
        MAE      & Bound & \textbf{0.0465} & 0.0480 & 0.0500 & 0.0498 & 0.0478 \\
                 & Core    & 0.0336 & 0.0350 & 0.0372 & 0.0364 & \textbf{0.0328} \\
        \addlinespace
        DE [\%]  & Bound & 6.41 & 5.80 & 5.47 & 5.10 & \textbf{5.08} \\
                 & Core    & 1.62 & 1.58 & 1.50 & 1.33 & \textbf{1.12} \\
        \midrule
        \multicolumn{7}{c}{\textbf{WGAN1e-3}} \\
        \midrule
        MRE [\%] & Bound & 39.55 & 38.78 & 38.33 & \textbf{38.10} & 38.30 \\
                 & Core    & 12.33 & 11.82 & 11.55 & \textbf{11.45} & 11.65 \\
        \addlinespace
        MAE      & Bound & 0.0465 & 0.0465 & 0.0450 & 0.0439 & \textbf{0.0429} \\
                 & Core    & 0.0336 & 0.0323 & 0.0316 & \textbf{0.0315} & 0.0322 \\
        \addlinespace
        DE [\%]  & Bound & 6.41 & 6.19 & 6.01 & 5.86 & \textbf{5.85} \\
                 & Core    & 1.62 & 1.61 & 1.57 & 1.48 & \textbf{1.38} \\
        \midrule
        \multicolumn{7}{c}{\textbf{WGAN1e-4}} \\
        \midrule
        MRE [\%] & Bound & 39.55 & 39.04 & 38.85 & \textbf{38.71} & 38.90 \\
                 & Core    & 12.33 & 11.96 & 11.77 & \textbf{11.70} & 11.84 \\
        \addlinespace
        MAE      & Bound & 0.0465 & 0.0451 & 0.0442 & \textbf{0.0436} & \textbf{0.0436} \\
                 & Core    & 0.0336 & 0.0324 & 0.0318 & \textbf{0.0315} & 0.0319 \\
        \addlinespace
        DE [\%]  & Bound & 6.41 & \textbf{6.30} & 6.31 & 6.40 & 6.65 \\
                 & Core    & 1.62 & 1.68 & 1.70 & 1.66 & \textbf{1.59} \\
        \bottomrule
        \end{tabular}
\end{table}

\subsection{Feature distribution analysis}

\begin{figure*}[ht]
  \centering
  \includegraphics[width=0.95\linewidth]{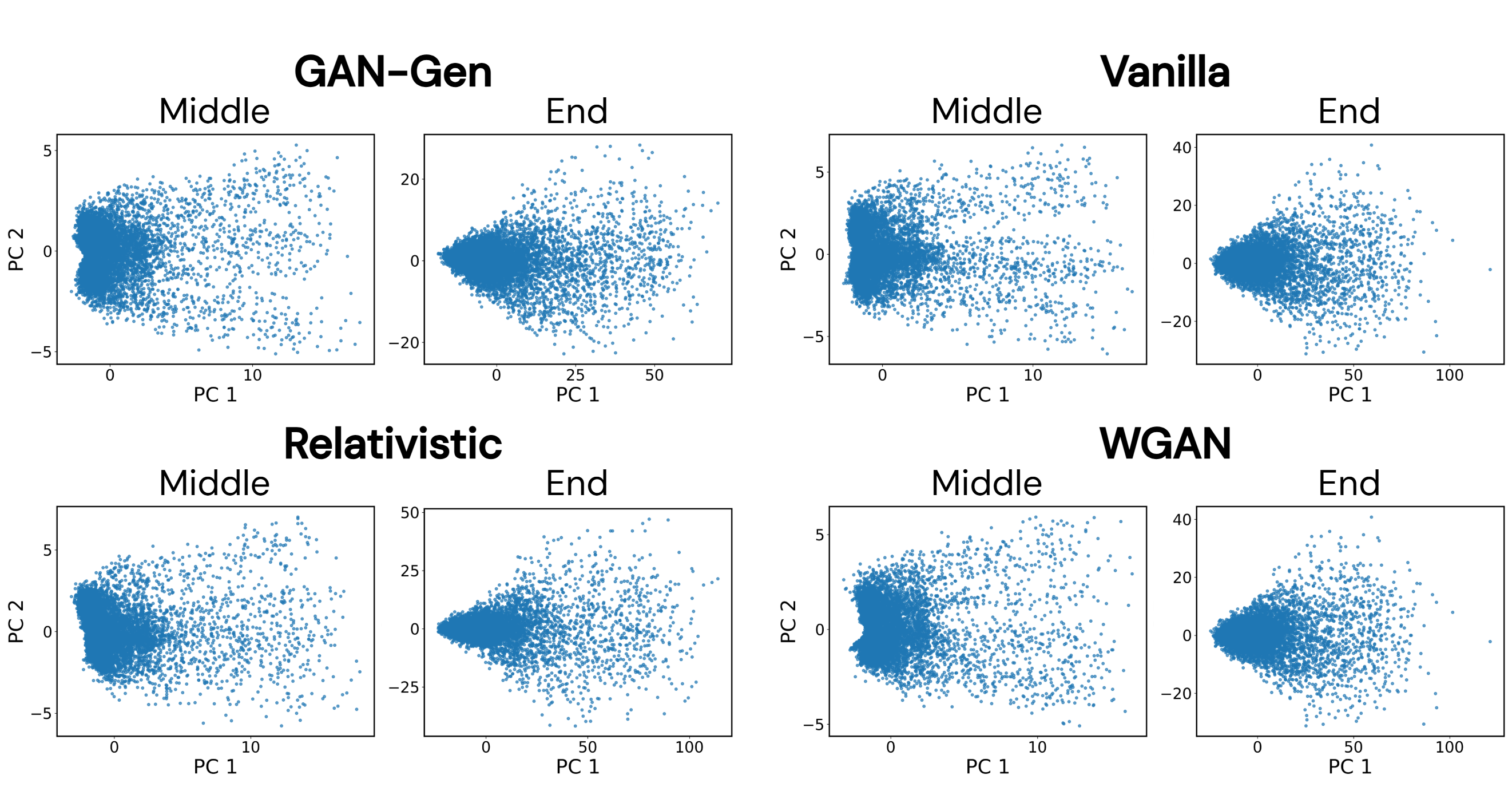}
  \caption{PCA projections of internal features for GAN-Gen, Vanilla, Relativistic, and WGAN networks, using 10,000 random samples of training data. Each subfigure depicts a 2D scatter plot of feature activations either before trilinear upsampling (middle) or prior to the final velocity direction branches (end). Despite differing training dynamics and performance, the feature distributions remain qualitatively similar.}
  \label{fig:Feature_distribution_vis}
\end{figure*}

Figure~\ref{fig:Feature_distribution_vis} shows two-dimensional Principal Component Analysis (PCA) projections of intermediate and near-final network features for GAN-Gen, Vanilla, Relativistic, and WGAN networks, respectively. Notably, despite the different adversarial training losses, the feature distributions exhibit visual similarities at both middle (pre-trilinear upsampling) and end layers (prior to velocity direction branching). At both positions, overall cluster shapes and densities are largely similar across all networks, suggesting that each network converges to comparable internal representations. 

\subsection{In-vivo evaluation}

\begin{figure*}[ht]
  \centering
  \includegraphics[width=1.0\linewidth]{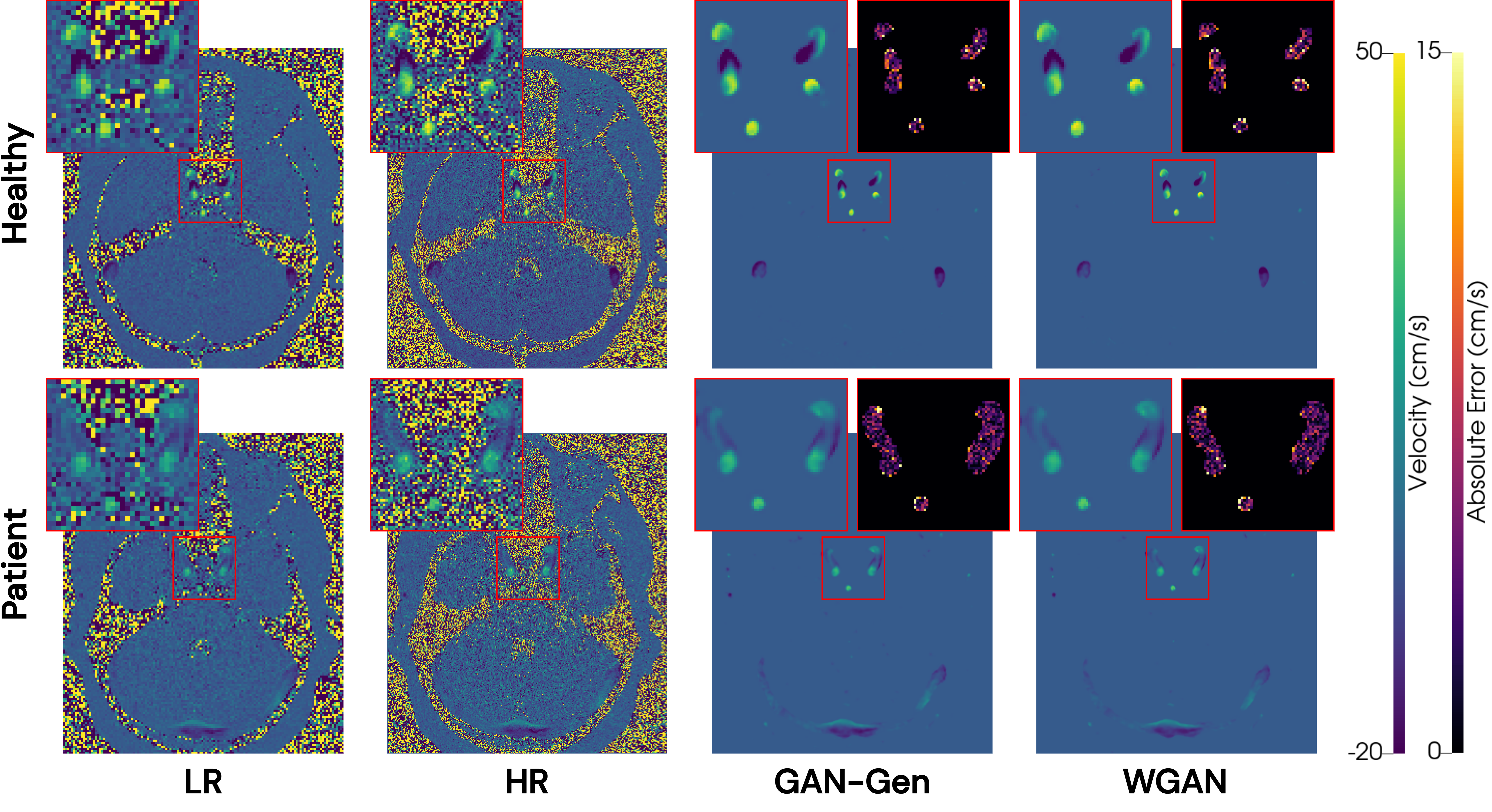}
  \caption{Qualitative comparison at peak systole between downsampled low-resolution (LR), native high-resolution (HR) \textit{in-vivo} acquisitions, and super-resolution (SR) reconstructions from GAN-Gen and WGAN on an axial slice of a healthy volunteer and a patient. Corresponding absolute error maps (HR vs. SR), masked using TOF segmentation, are also shown.}
  \label{fig:Invivo}
\end{figure*}

\begin{table*}[!b]
\centering
\small
\caption{Average performance metrics for all networks across the fluid region of the \textit{in-vivo} datasets, computed over the full cardiac cycle. The best values are highlighted in bold.}
\label{tab:multirow}
\begin{tabular}{l l c c c c c c}
\toprule
\textbf{Case} & \textbf{Model} & $\, \,$ MRE [\%] & MAE & vNRMSE
& $\, \,$ DE [\%] & \(k\) & \(R^2\) \\
\midrule
\multirow{5}{*}{\textbf{Healthy}} 
 & 4DFlowNet
   & 37.93  & 0.1586 & 0.1114 & 7.20 
   & (0.734, 0.732, 0.689)
   & (0.791, 0.773, 0.736) \\
 & GAN-Gen
   & \textbf{27.24}  & \textbf{0.1054} & \textbf{0.0787} & \textbf{3.67} 
   & (\textbf{0.900}, \textbf{0.885}, \textbf{0.843})
   & (\textbf{0.886}, \textbf{0.858}, \textbf{0.858}) \\
 & Vanilla
   & 28.23  & 0.1105 & 0.0822 & 3.98 
   & (0.890, 0.852, 0.827)
   & (0.877, 0.849, 0.845) \\
 & Relativistic
   & 28.06  & 0.1110 & 0.0827 & 3.88 
   & (0.840, 0.781, 0.812)
   & (0.877, 0.851, 0.845) \\
 & WGAN
   & 27.57  & 0.1072 & 0.0803 & 3.82 
   & (0.886, 0.863, 0.841)
   & (0.884, 0.854, 0.852) \\
   \midrule
\multirow{5}{*}{\textbf{Patient (MS)}} 
 & 4DFlowNet
   & 40.34  & 0.1079 & 0.0795 & 8.19 
   & (0.734, 0.727, 0.622)
   & (0.794, 0.749, 0.647) \\
 & GAN-Gen
   & \textbf{32.13}  & \textbf{0.0811} & \textbf{0.0638} & \textbf{5.51} 
   & (\textbf{0.873}, \textbf{0.847}, 0.738)
   & (\textbf{0.862}, \textbf{0.832}, 0.743) \\
 & Vanilla
   & 32.73  & 0.0832 & 0.0651 & 5.62 
   & (0.859, 0.812, 0.738)
   & (0.858, 0.828, 0.737) \\
 & Relativistic
   & 32.87  & 0.0841 & 0.0654 & 5.72 
   & (0.806, 0.762, 0.722)
   & (0.859, 0.829, \textbf{0.744}) \\
 & WGAN
   & 32.50  & 0.0825 & 0.0647 & 5.58 
   & (0.857, 0.830, \textbf{0.745})
   & (0.861, 0.827, 0.741) \\
\bottomrule
\end{tabular}
\label{table:invivo_metrics}
\end{table*}

Qualitative visualization of SR outputs from GAN-Gen and WGAN compared against the \textit{in-vivo} HR reference is shown in Figure~\ref{fig:Invivo}. Visually, the network enhances fluid regions while simultaneously eliminating noise in surrounding static tissue. Consistent with the qualitative observations, the GAN-based approaches achieved strong performance in both the healthy (vNRMSE: 0.1054–0.1072) and patient (vNRMSE: 0.0638–0.0654) cases, whereas 4DFlowNet showed comparatively poorer accuracy (vNRMSE healthy: 0.1114, patient: 0.0795). In line with the \textit{in-silico} findings, GAN-Gen and WGAN outperformed the Vanilla and Relativistic variants across most metrics, although the improvements were modest. GAN-Gen obtained the best overall scores (18 out of 20 metrics), yet the differences relative to WGAN were marginal and should be interpreted with caution given the uncertainty of the \textit{in-vivo} HR reference.

\section{Discussion}

In this study, we have investigated the use of GANs for super-resolving and denoising 4D Flow MRI data, with particular emphasis on near-wall flow recovery in a cerebrovascular setting. With a dedicated GAN architecture implemented and three adversarial loss formulations evaluated, the Wasserstein implementation demonstrated superior training stability and output accuracy as compared to conventional setups. Further, the Wasserstein GAN also enabled slight improvements in boundary flow recovery relative to a generator-only baseline, particularly under high-noise conditions. However, the overall performance gains were modest, underscoring the challenges of effectively leveraging adversarial losses in this context. Nonetheless, these findings highlight the potential of adversarial learning for enhancing super-resolution 4D Flow MRI.

\subsection{Quantitative results}

Comparison between the baseline residual network 4DFlowNet~\cite{Ferdian2023Cerebrovascular} and our generator-only setup (GAN-Gen) demonstrated improved accuracy when using the generator-converted approach (Table~\ref{table:overall_metrics}). Since both approaches were trained without any adversarial loss, these improvements can instead be attributed to minor differences in architecture (GAN-Gen utilizing residual-in-residual dense blocks rather than conventional residual  blocks~\cite{wang2018esrgan}), as well as differences in data input. For the latter, we opted for utilizing velocity data as the sole input whereas 4DFlowNet incorporated both velocity \emph{and} magnitude information. While additional input channels could provide richer information, preliminary testing indicated that including magnitude data introduced noise into the learning process. Although possibly beneficial at low SNR settings (Tables~\ref{Table:lowSNR_combined}) where magnitude information can assist the network in identifying velocity data from background signal, overall performance was empirically degraded. 

Moving on to the impact of introduced adversarial entities, training instabilities were observed for both Vanilla and Relativistic setups with a negative impact on recovered accuracy as compared to both the GAN-Gen baseline and the baseline 4DFlowNet (Table~\ref{table:overall_metrics}). In contrast, the shift to a Wasserstein GAN (WGAN) setup showed improved stability with slightly improved accuracy in near-wall regions. This superior boundary performance likely stems from the adversarial loss’s ability to enforce sharper and more realistic spatial transitions~\cite{ledig2017srgan}, which are critical for capturing steep velocity gradients. By comparison, GAN-Gen achieved lower errors in the vessel core, where blood flow patterns transition in a smoother pattern, and where voxel-wise MSE can be particularly effective at velocity reconstruction without adversarial artifacts.

Our results also revealed varying performance as a function of SNR, with WGAN outperforming the generator-only model (GAN-Gen) on input data with comparably low SNR, whereas GAN-Gen achieved superior performance under contrasting high SNR conditions (Tables~\ref{Table:highSNR_combined} and~\ref{Table:lowSNR_combined}). In low SNR settings, relying solely on a voxel-wise loss can become unreliable as high measurement noise makes it difficult to infer meaningful signals at a single-voxel level. By contrast, the adversarial framework pushes the generator to synthesize realistic velocity fields, guiding it to capture the underlying flow structures rather than merely matching data at a voxel-level scale~\cite{xie2018tempogan, kim2021unsupervised}. Conversely, at high SNR, the underlying signal is better preserved even at a voxel-level scale, benefiting the GAN-Gen setup which is void of possible adversarially-driven artifacts that can arise~\cite{sajjadi2017enhancenet, liang2022details}. As such, the optimal choice of loss may depend on underlying data quality, where adversarial setups may offer advantages in settings of exaggerated noise or data imperfections. Taken together, these findings indicate that adversarial learning is not universally superior to voxel-wise supervision in this setting, but rather provides conditional benefits.

Consistent with the CFD-based evaluation, the models also demonstrated good performance on the \textit{in-vivo} datasets (Tables~\ref{table:invivo_metrics}), indicating that the training setup - and in particular the dedicated downsampling pipeline - captures the characteristics of acquired data. The comparatively poor performance of 4DFlowNet, in addition to architectural differences, likely reflects inconsistencies between co-registered and \textit{in-vivo} magnitude images. GAN-Gen achieved the best performance metrics, but it remains difficult to disentangle improvements in flow recovery from the influence of measurement noise in the HR reference. This challenge is especially pronounced for boundary flow recovery, where segmentation uncertainties, partial voluming, and noise complicate voxel-level comparisons near vessel walls. Beyond voxel-wise metrics, future validation could incorporate physics-based consistency measures, such as flow-rate conservation, WSS continuity, or divergence analysis. These approaches may provide indirect evidence of physically coherent solutions in the absence of ground-truth reference data and thus represent a promising direction for strengthening clinical validation.

\subsection{Vanilla and Relativistic GAN instabilities}

%\textit{Analysing Figure 4 - training instabilities}
As extensively reported in the literature~\cite{arjovsky2017wasserstein, mescheder2018training, saxena2021review, gui2021review, karnewar2020msg, xu2020understanding}, the adversarial nature of the GAN setup comes with characteristic training instabilities. These tendencies are also confirmed in our work, where the introduction of both an out-of-the-box vanilla implementation as well as a relativistic counterpart leads to distinct performance degradation (seen in e.g., the loss curve tendencies in Figure~\ref{fig:Training_curves_1}.A). Elucidating these effects, Figure~\ref{fig:Training_curves2} provides a more detailed view of the adversarial loss components: when no adversarial feedback was applied to the generator ($\lambda_{\text{G}}=0$; first column of Figure~\ref{fig:Training_curves2}) all variants exhibited expected convergent behavior with the discriminator trained independently from any adversarial input. Once adversarial input was switched on, however, substantial differences emerged (second column). For the Vanilla and Relativistic GANs the discriminator exhibited increasing difficulties in separating HR and SR input. A large discrepancy appeared between training and validation discriminator losses, suggesting overfitting of the discriminator to the training distribution and poor generalization to unseen data. Further, the generator adversarial losses (third column) exhibited a steady increase over time, reflecting the generator’s inability to effectively adapt and improve its outputs under this adversarial training. 

Connecting to published work, several possible explanations exist for the observed differences in training stability. Although Vanilla and Relativistic GANs have demonstrated strong performance for single-image super-resolution (SISR) tasks~\cite{ledig2017srgan, wang2018esrgan}, these formulations are known to suffer from inherent training instabilities~\cite{mescheder2018training, karnewar2020msg}. For the Vanilla setup, Arjovsky et al.~\cite{arjovsky2017wasserstein} highlight how instabilities may arise from the inherent design of minimizing the Jensen-Shannon (JS) divergence between real and generated data distributions; a setup that - when the two distributions exhibit little or no data overlap - may result in flattened divergence and vanishing gradients, preventing the generator from receiving meaningful training updates. Likewise, other possible explanations for the Vanilla behavior may lie in the lack of gradient regularization and continuity constraints, where irregular gradients, non-convex loss landscapes, and noisy updates destabilize effective training. For the Relativistic GAN, even though the setup is conceptually altered (the discriminator objective being to predict the \textit{probability} that a real sample is more realistic than a generated one~\cite{jolicoeur2018relativistic, wang2018esrgan}), training remains fundamentally based on a similar adversarial minimax game as in the original Vanilla setup. As such, the aforementioned underlying stability issues (non-informative gradients, discriminator dominance) remain inherently unresolved. 

Herein, it is important to note that while the Vanilla and Relativistic GANs may exhibit unstable training dynamics due to inherent mathematical setups as per above, outcomes may also be partially attributed to an imbalance between generator and discriminator capacities. As reported in prior work~\cite{heusel2017gans}, the relative strength of these two networks plays a critical role in shaping GAN training behavior. An overly powerful discriminator can easily distinguish real from generated data, leading to vanishing gradients for the generator. Conversely, an underpowered discriminator may fail to provide meaningful gradients to the generator. In this context, architectural design choices such as depth, filter width, and residual connectivity, as well as the inclusion of regularization techniques, may directly impact observed adversarial equilibrium. To exemplify, in preliminary experiments we observed that omitting or underweighting the L2-regularization term of the discriminator led to overfitting (evidenced by sharp divergence between training and validation losses), whereas setting the regularization too high resulted in opposite underfitting. Such and other optimization-related hyperparameters could offer further levers for improving stability, however, they still highlight the sensitivity of these implementations.

\subsection{Wasserstein GAN convergence and achieving training stability}

In contrast to the aforementioned instability issues, switching to a WGAN implementation demonstrated more favorable training dynamics. As highlighted in Figure~\ref{fig:Training_curves2} (second column), using the WGAN setup the discriminator starts to separate HR from SR when adversarial loss is introduced. Concurrently, the generator adversarial loss (third column) initially increases before stabilizing, indicating that the generator is able to adapt to the discriminator feedback to a degree not observed in the Vanilla or Relativistic loss implementations. 

To understand the benefits of the WGAN implementation, it is imperative to outline the two key ways in which the implementation modifies the adversarial task: (1) the discriminator is trained to approximate the Earth Mover (Wasserstein-1) distance between the real and generated data distributions, providing a continuous and meaningful training signal even when the distributions have limited overlap~\cite{arjovsky2017wasserstein}; and (2) a gradient penalty is introduced to enforce the 1-Lipschitz constraint on the discriminator~\cite{gulrajani2017improvedwgan}, promoting smoother gradients and stabilizing training dynamics. Beyond theoretical grounding, WGAN has demonstrated performance in SISR, improving both SR performance and training stability across architectures and hyperparameters~\cite{chen2017face, tang2022single}. Our findings are in agreement with these reports, as the WGAN exhibited robust performance \textit{without} requiring extensive hyperparameter tuning - of clear benefit in our specific super-resolution setting. This practical advantage, suggests that WGAN may provide a more reliable and accessible framework for future work, not least applying generative models in our specific use-case of enhanced 4D Flow MRI.

While the WGAN configuration demonstrated improved training stability and marginally better boundary performance relative to the generator-only baseline, the overall perceptual differences in super-resolved outputs remained modest (see Figure~\ref{fig:interpolation_vis}). Unlike prior SISR works such as ESRGAN~\cite{wang2018esrgan}, where adversarial training introduces pronounced improvements in edge sharpness and contrast, our results showed incremental contrast variations in performed weight interpolation experiments (Figure~\ref{fig:interpolation_vis}). Interpolated models between the generator-only and fully adversarial WGAN setups yielded visually similar outputs, with improvements difficult to disentangle from continued generator training alone. A potential explanation for this outcome likely reflects fundamental differences between the SISR task and 4D Flow MRI velocity field super-resolution. In classical SISR, ground truth HR images often contain high-frequency details and rich edge structures, allowing the discriminator to guide the generator toward sharper, more textured outputs. In contrast, in our setting of realistically sampled 4D Flow MRI data, high-resolution data still represent a comparably coarse representation of the fine, intracranial vessel domain. Further, the downsampling procedure of a clinical MR system (as per Figure~\ref{fig:DV-recon}) introduces non-trivial modifications to the higher-resolution data, where high-frequency detail may be convolved in a non-trivial way, as compared to a natural image case. Another possible explanation lies in the fundamentally different structure of the underlying data. Whereas images contain high-frequency components and texture patterns, fluid velocity fields are governed by the Navier–Stokes equations and exhibit strong spatial smoothness and continuity constraints. This may inherently limit the admissible solution space for the discriminator, effectively constraining the extent to which adversarial learning can introduce additional high-frequency structure without violating physical plausibility. While this interpretation is hypothetical and requires further systematic investigation, it suggests that the role of adversarial learning in physics-constrained domains may differ fundamentally from that in conventional computer vision tasks.

\subsection{Scientific contextualization}
Since it's conception~\cite{goodfellow2014gan}, the extension of GANs into the domain of SISR has seen comparably widespread utility with a variety of application areas addressed \cite{ledig2017srgan, wang2018esrgan, sajjadi2017enhancenet, karnewar2020msg}. GAN-based SISR has also seen growing adoption in the setting of medical imaging, not least with a few notable MRI-based examples. In 2022, Zhang et al. presented the SOUP-GAN approach, achieving super-resolution conversion of T1-weighted brain MRI using a Vanilla-based implementation without reporting any instability concerns during training~\cite{zhang2022soup}. In the same intracranial setting, Ran et al. explored a Wasserstein implementation through RED-WGAN~\cite{ran2019denoising}, and a variety of other examples exist exploring the Wasserstein loss as a key for achieving training stability~\cite{jiang2019accelerating, zhao2022generative}. To the best of our knowledge, our work represents the first GAN implementation for 4D Flow MRI. However, for lower-order imaging Morales et al.~\cite{morales2025accelerated} recently introduced the GAN-based CRISPFlow network, achieving super-resolution of 2D phase-contrast MRI. In line with our findings, their approach improved image sharpness and flow fidelity. However, in contrast to our work, near-flow velocity enhancement was not analysed separately, nor were data of higher-order dimensions assessed. 

Beyond the GAN-setting, exploration of super-resolution 4D Flow MRI has so far been divided into convolutional~\cite{Rutkowski2021, shit2022srflow, Ferdian20204DFlowNet, patel2025super} or coordinate-based setups~\cite{kissas2020machine, fathi2020super, shone2023deep, saitta2024implicit}. In particular, the latter includes exploration of so called physics-informed neural networks (PINN), where governing fluid mechanical principles are included in the loss function by ensuring that the converged solution abides to fundamental hemodynamic laws. While highly efficient in predicting both velocity and pressure data at virtually arbitrary spatiotemporal sampling, the patient-specific nature of these models requires re-training for each new patient, limiting clinical translation~\cite{shone2023deep}. Promising novel variations are here emerging, including unsupervised~\cite{sautory2024unsupervised, gormezano2024denoising} or implicit neural representations~\cite{saitta2024implicit}, however, these remain to be explored in more MR-like or patient-specific datasets. Overall, head-to-head comparisons between presented networks (including our GAN-variations) would be of direct interest to the community, not least with regard to near-wall velocity recovery. However, its realization remains non-trivial with key alternative networks not yet publicly available, or with remaining implementations needed to extend these into the specific data setting of 4D Flow MRI.

\subsection{Clinical contextualization}

The clinical applicability of our work can be linked to (\textit{i}) the accuracy of the underlying CFD simulations, (\textit{ii}) the realism of the synthetic image generation, and (\textit{iii}) the relevance of super-resolution 4D Flow MRI. 

Regarding the first point, we sought to utilize high-fidelity simulations as training data to enable precise, and highly regional interrogation of network quality in a setting void of real-life sources of error such as acquisition noise or movement artifacts. By using input data from a highly refined and validated image-based CFD framework~\cite{schollenberger2021} we ensure translational validity, with simulated input data accurately representing typical hemodynamic settings encountered in an intracranial setting.

For the second point, we sought to transfer the CFD data into a realistic image setting by closely mimicking the sub-sampling process in an actual MRI scanner. Although this approach does not account for \textit{all} possible scanner or patient-specific factors, such as spatially varying coil sensitivities, patient-specific motion artifacts, or underlying MR physics~\cite{dirix2022synthesis}, it seeks to include \textit{essential} aspects of the noise characteristics encountered in real MRI acquisitions, not least considering the specific value of dual-venc acquisitions highlighted in an intracranial setting~\cite{Schnell2017}. The strong performance observed on the \textit{in-vivo} dual-venc acquisitions suggests that this strategy captures the characteristics relevant for clinical translation. Notably, while our approach is tailored to this dual-venc setup, our framework is not inherently restricted to this sequence type. Instead, by systematically varying SNR levels and venc settings in the synthetic training data, we sought to establish a flexible training paradigm applicable and extendable to other acquisition sequences as well. An additional consideration relates to the use of magnitude images derived exclusively from healthy in-vivo acquisitions. While the use of in-vivo magnitude images improves the realism of resulting velocity noise patterns - by incorporating authentic coil sensitivities and tissue contrasts - it does not explicitly model pathological magnitude alterations. Incorporating such pathology-specific magnitude features - e.g., signal voids from intravoxel dephasing or heterogeneous plaque-related intensities - remains challenging, partly due to the difficulty of accurately co-registering pathology-specific magnitude characteristics with anatomically matched regions in the CFD-derived velocity fields. As a result, the synthetic training data may underrepresent magnitude-domain characteristics encountered in complex vascular disease. Although the favorable in-vivo translation suggests a degree of robustness to this domain gap, future work could explore incorporation of pathological magnitude datasets together with targeted validation in disease-specific cohorts to further assess generalization under such conditions.

For the third point, the clinical value of 4D Flow MRI lies in accessing functional hemodynamic metrics across various cardiovascular compartments in a purely non-invasive fashion. While used extensively to study large-vessel flows~\cite{catapano20204d, demirkiran2022clinical}, the technique has limitations when extended into narrower vessel compartments with estimation biases stemming primarily from limited spatiotemporal recovery~\cite{aristova2019standardized, marlevi2021noninvasive, knapp2023fetal, hyodo20224d}. In this sense, our study focuses on extending the technique's capabilities into small-vessel domains, \textit{specifically} targeting near-wall flows which are related to wall shear stress (WSS), oscillatory shear index, or endothelial activation; all entities shown to be directly related to various cardiovascular disease forms~\cite{szajer2018comparison, yamada2021quantification, takehara20204d}. Our adversarial approach provides encouraging improvements in this regard. However, given the modest magnitude of these gains, further work is needed to determine whether such improvements translate into measurable benefits in a clinical setting. Regardless, our work seeks to outline the potential and challenges of alternative generative approaches in the setting of super-resolution imaging.

\subsection{Limitations}

A few limitations are worth pointing out. First, we employed a limited number of cerebrovascular models (n=4) in our training and validation framework. While the inclusion of more models could enhance generalization, the patch-based training approach is specifically designed to limit dependence on global geometry and extend application. Recent work has also explored cross-compartment generalization strategies to improve super-resolution performance across diverse vascular domains~\cite{ericsson2024generalized}, however, this remains to be explored in the setting of generative models. Second, although focused on the super-resolution capabilities, it should be noted that our proposed networks are tasked with multiple objectives: segmenting fluid regions, denoising, and super-resolving. While this design offers practical simplicity and end-to-end learning, it may also impose competing demands on the network’s capacity, possibly limiting the ability to optimally solve each task. Dedicated approaches - where separate models handle segmentation, denoising, and super-resolution independently - might improve overall performance, however, this remains to be assessed. Third, while the regional decomposition of the loss function into non-fluid, boundary and core regions was motivated by the clinical importance of near-wall velocity recovery, we did not perform a dedicated ablation study to isolate its independent contribution relative to other loss formulations. Similarly, alternative architectural designs and complexity levels may influence both reconstruction accuracy and adversarial training dynamics. A systematic ablation study disentangling such factors could provide additional insight into optimal design choices and is left for future investigation.

Lastly, although significant efforts were made to enhance the clinical relevance of our synthetic 4D Flow data, and preliminary evaluation on two \textit{in-vivo} datasets (one healthy volunteer and one patient with multiple sclerosis) was included, the limited scope of this testing restricts the conclusions that can be drawn. In particular, reliable assessment of boundary recovery remains challenging due to segmentation uncertainty, partial voluming, and noise. Furthermore, the small sample size and the absence of cases with diverse cerebrovascular pathologies (e.g., aneurysms, stenoses, or tortuous segments) limit the ability to assess generalization. Future work should therefore expand \textit{in-vivo} validation across larger and more heterogeneous patient cohorts, and explore hybrid strategies combining CFD-based and clinical data or transfer-learning approaches to better bridge the gap between synthetic training and clinical applicability.

\section{Conclusion}

In this study, we present a GAN-based framework for super-resolving and denoising 4D Flow MRI data with a particular focus on cerebrovascular near-wall flow. Through systematic evaluation on synthetic datasets, we demonstrate both the promise and the challenges of adversarial learning, specifically showing that a Wasserstein implementation achieves training stability and incremental improvement on near-wall velocity recovery compared to non-adversarial, voxel-wise networks. These findings underscore the potential and current limitations of adversarial learning in advancing 4D Flow MRI, and provide a foundation for future work exploring the use of generative networks, including diffusion-based models, for improved hemodynamic quantification in a clinical setting.

\section{Acknowledgements}

Computations were performed on resources provided by the National Academic Infrastructure for Supercomputing in Sweden at the National Supercomputer Centre at Linköping University (Berzelius). This work was supported in part by European Union ERC, MultiPRESS, under Grant 101075494. The views and opinions expressed are those of the authors and do not reflect those of the European Union or the European Research Council Executive Agency.

\section{Declaration of generative AI and AI-assisted technologies in the writing process}
During the preparation of this work, the authors used ChatGPT (OpenAI) to assist in improving the readability and language of the manuscript. After using this tool, the authors reviewed and edited the content as needed and take full responsibility for the content of the published article.

 \bibliographystyle{elsarticle-num} 
 \bibliography{references}

\end{document}